\documentclass[journal]{IEEEtran}

\usepackage{xcolor,soul,framed} 

\colorlet{shadecolor}{yellow}
\usepackage[pdftex]{graphicx}
\graphicspath{{../pdf/}{../jpeg/}}
\DeclareGraphicsExtensions{.pdf,.jpeg,.png}

\usepackage[cmex10]{amsmath}
\usepackage{array}
\usepackage{mdwmath}
\usepackage{mdwtab}
\usepackage{eqparbox}
\usepackage{url}
\usepackage[colorlinks,
            linkcolor = blue,
            anchorcolor = blue,
            citecolor = blue,
            urlcolor = blue]{hyperref}
\begin{document}
\bstctlcite{IEEEexample:BSTcontrol}
    \title{Gravitational Communication: Fundamentals, State-of-the-Art and Future Vision}
  \author{Houtianfu Wang,~\IEEEmembership{Student Member,~IEEE,}
      Ozgur B. Akan,~\IEEEmembership{Fellow,~IEEE}\\

\thanks{The authors are with Internet of Everything Group, Department of Engineering, University of Cambridge, CB3 0FA Cambridge, UK.}
  \thanks{Ozgur B. Akan is also with the Center for neXt-generation Communications
(CXC), Department of Electrical and Electronics Engineering, Koç University,
34450 Istanbul, Turkey (email:oba21@cam.ac.uk)}
  }


\maketitle

\begin{abstract}

This paper provides a comprehensive overview of fundamentals and the latest research progress in gravitational communication, with a detailed historical review of gravitational wave generation and detection. Key aspects covered include the evolution of detection sensitivity and generation methods, modulation techniques, and gravitational communication channel. While gravitational wave communication holds promise for overcoming limitations in traditional electromagnetic communication, significant challenges remain, particularly in wave generation and detection. The paper also explores various modulation techniques and examines environmental influences on gravitational wave transmission. A comparative discussion is provided between gravitational and classical communication modalities—including electromagnetic, quantum, particle, acoustic, and optical communications—highlighting the strengths and limitations of each. Furthermore, potential application and future vision for gravitational communication are also envisioned. Finally, Potential research directions to bridge the gap between theoretical and practical applications of gravitational wave communication are investigated.

\end{abstract}


%
\IEEEpeerreviewmaketitle

\section{Introduction}

\IEEEPARstart{G}{ravitational} waves are ripples in the fabric of spacetime generated by the acceleration of massive objects, such as merging black holes or neutron stars. Predicted by Albert Einstein's general theory of relativity in 1916 \cite{einstein1916}, these waves propagate at the speed of light, carrying information about their cataclysmic origins and the nature of gravity itself. The first direct detection of gravitational waves in 2015 by the LIGO \cite{abramovici1992} and Virgo \cite{caron1997} collaborations was a milestone in physics and astronomy, confirming Einstein's century-old prediction and inaugurating a new era of gravitational wave astronomy.

The discovery of gravitational waves has opened a new observational window for astronomy and physics, offering a unique approach to exploring the depths of the universe and extreme astrophysical phenomena. Beyond its impact on astronomical research, gravitational waves have also garnered widespread attention as a new communication paradigm \cite{baker2012}. Gravitational communication, also known as gravitational wave communication, holds the promise of overcoming the limitations of traditional electromagnetic communication, enabling robust transmission across extreme environments and vast distances \cite{ye2022}. 

Currently, research on gravitational communication focuses primarily on key technologies such as the generation, detection, modulation, and propagation of gravitational waves. Generating gravitational waves under laboratory conditions is a critical challenge to advance gravitational communication. Researchers have explored various innovative methods to achieve this, including mechanical resonance and rotational devices \cite{weber1960}, superconducting materials \cite{petlan2001}, and particle beam collisions \cite{baker2004}, as well as techniques involving high-power lasers and electromagnetic fields \cite{pustovoit2021}, \cite{morozov2021}.  

In terms of gravitational wave detection, the application of deep learning and artificial intelligence is revolutionizing signal processing methods, improving the accuracy of gravitational wave detection in complex noise environments. Additionally, indirect detection methods that utilize the interaction between gravitational waves and electromagnetic fields, magnons, or photons offer a feasible pathway to realizing gravitational communication. These methods not only overcome the sensitivity limitations of direct detection but also provide new technical means for information transmission and retrieval. 

For the modulation of gravitational waves, researchers have proposed various techniques, such as amplitude modulation through astrophysical phenomena \cite{qiu2022}, frequency modulation involving dark matter \cite{wang2023}, manipulation of gravitational waves using superconductors \cite{woods2005}, and modulation based on non-metric theories of gravity \cite{babourova2018}. These modulation techniques provide the possibility of encoding and transmitting information via gravitational waves. However, gravitational waves may be affected by cosmic perturbations, radiation, electromagnetic fields and other environmental factors during their propagation, which can change the characteristics of the channel and lead to signal effect, thus affecting the quality and reliability of communication. Therefore, when designing a gravitational communication system, these environmental effects deserve full consideration and the communication channel must be thoroughly analyzed to ensure reliable signal transmission in the complex cosmic environment.

This paper aims to systematically explore the fundamentals and the latest research developments in gravitational communication, analyze current technical challenges, and explore future directions. In this direction, section II provides the explanation of physical foundations of universal gravity theory and its implications on communications. Subsequently, section III explores various methods for generating gravitational waves, highlighting mechanical resonance, superconducting techniques, and high-power laser approaches. Section IV discusses gravitational wave detection methods, emphasizing direct laser interferometry, deep learning advancements, and indirect detection approaches. Section V explores gravitational wave modulation techniques, including amplitude modulation from astrophysical phenomena, frequency modulation influenced by dark matter, and superconducting-based methods. Section VI examines the gravitational wave communication channel components, analyzing propagation effects like frequency shifts, attenuation, polarization changes, phase distortion, and fading, along with their implications for system design. Section VII presents the comparative strengths and limitations of gravitational wave communication versus classical communication methods. Section VIII explores potential applications of gravitational wave communication in extreme environments and deep-space scenarios, envisioning its future role in advancing information transmission technologies. Finally, section IX concludes the paper by recaps main findings and suggesting directions for future research.

\begin{figure*}[!t]
    \centering
    \includegraphics[width=\linewidth]{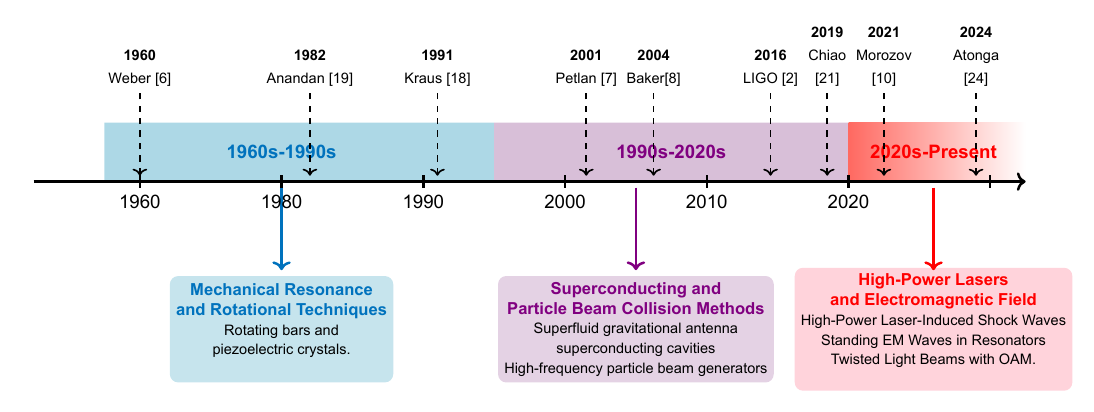}
    \caption{Timeline of key developments in gravitational wave technology and detection methods.}
    \label{fig:timeline}
\end{figure*}

\section{Fundamentals of Gravity and Gravitational Communication}

Before delving into the technical details of gravitational communication, it is worthwhile to revisit the underlying physical principles that govern these processes and to clarify the motivations driving research in this emerging field. Gravitational communication rests on the cornerstone of universal gravity theory and its modern extension—general relativity. While gravitational concepts date back to Newton’s formulation of universal gravitation, it was Einstein’s revolutionary description of gravity as the curvature of spacetime that truly set the stage for harnessing gravitational waves as a potential information carrier. This section provides an overview of the physical foundations of gravity and explores how they lay the groundwork for novel communication paradigms.


\subsection{Physical Foundations of Universal Gravity}
Historically, Newton’s law of universal gravitation posits that every mass exerts an attractive force on every other mass, proportional to the product of their masses and inversely proportional to the square of the distance between them \cite{newton2022}. Despite its impressive success in describing planetary motions and most terrestrial phenomena, Newtonian gravity has inherent limitations—most notably, it treats the gravitational influence as acting instantaneously across any distance, and it does not incorporate relativistic effects crucial for extreme conditions such as near-light-speed motion or strong gravitational fields.

Einstein’s general relativity (GR) recasts gravity not as a force but as a manifestation of curved spacetime \cite{einstein1916}. Matter and energy tell spacetime how to curve, while curved spacetime tells matter how to move. Within this geometric framework, orbits and trajectories can be viewed as objects following the “straightest possible paths” (geodesics) in a curved spacetime manifold. Critically, GR predicts that disturbances in this curvature—particularly from massive accelerating objects—will produce ripples that propagate at the speed of light, known as gravitational waves.

Gravitational waves (GWs) are remarkably weakly interacting with matter, passing through stellar interiors or dense media with negligible absorption or scattering \cite{einstein1916}. This transparency contrasts starkly with electromagnetic (EM) waves, which are readily blocked, scattered, or attenuated by plasma, dust, or atmospheric conditions. The small coupling to matter implies that detecting GWs demands incredibly sensitive instruments (e.g., laser interferometers) to measure minute changes in length scales caused by passing waves. Once measured, however, these faint signals can encode direct information about astrophysical events (like black hole or neutron star mergers) and, in principle, may serve as carriers of artificially modulated information.
\subsection{Implications for Communication} 

The unique propagation characteristics of gravitational waves open up possibilities for communication in environments hostile to conventional EM-based systems. In principle, GWs can penetrate dense materials, circumvent electromagnetic shielding, and traverse great cosmic distances without substantial loss of signal integrity. These attributes provide a tantalizing avenue for deep-space or subsurface communication, where EM signals could be severely attenuated or drowned out by background noise.

Beyond its practical allure, gravitational communication probes the very frontier of modern physics. Investigating whether and how information could be reliably transmitted via gravity can shed light on unresolved questions in quantum gravity, the nature of spacetime, and the limits of field quantization. If successful, even partial demonstrations of engineered gravitational signals would have profound implications, potentially reshaping our understanding of fundamental interactions and expanding our communication toolkit beyond traditional EM-based technologies.

Although the theoretical foundations of gravitational communication provide solid support for its potential applications, the key to transforming the concept into a practical means of communication lies in the ability to efficiently generate gravitational waves under controlled conditions. The following sections will explore the development history and major breakthroughs of these generation techniques in detail.

\section{Gravitational Wave Generation}

The generation of gravitational waves is pivotal for advancing gravitational communication, yet it remains one of the foremost challenges in contemporary technological development. Traditionally, gravitational waves are understood to emanate from the violent motions of massive celestial bodies, such as black hole mergers and neutron star collisions \cite{abbott2016}, \cite{abbott2017}. These macroscopic phenomena cannot be replicated under laboratory conditions, which limits the practical application of gravitational waves in communication. Consequently, researchers have been exploring innovative methods to generate gravitational waves under controlled conditions. This has led to the development of various techniques, each leveraging different principles and technologies. Fig.~\ref{fig:timeline} illustrates key developments in gravitational wave technology, highlighting major breakthroughs and the evolution of experimental approaches over the years.

\subsection{Mechanical Resonance and Rotational Techniques}

\subsubsection{Rotating Mass} One of the earliest approaches to generating gravitational waves involved mechanical devices that rely on rotating masses.

\begin{figure}[!t]
    \centering
    \includegraphics[width=\linewidth]{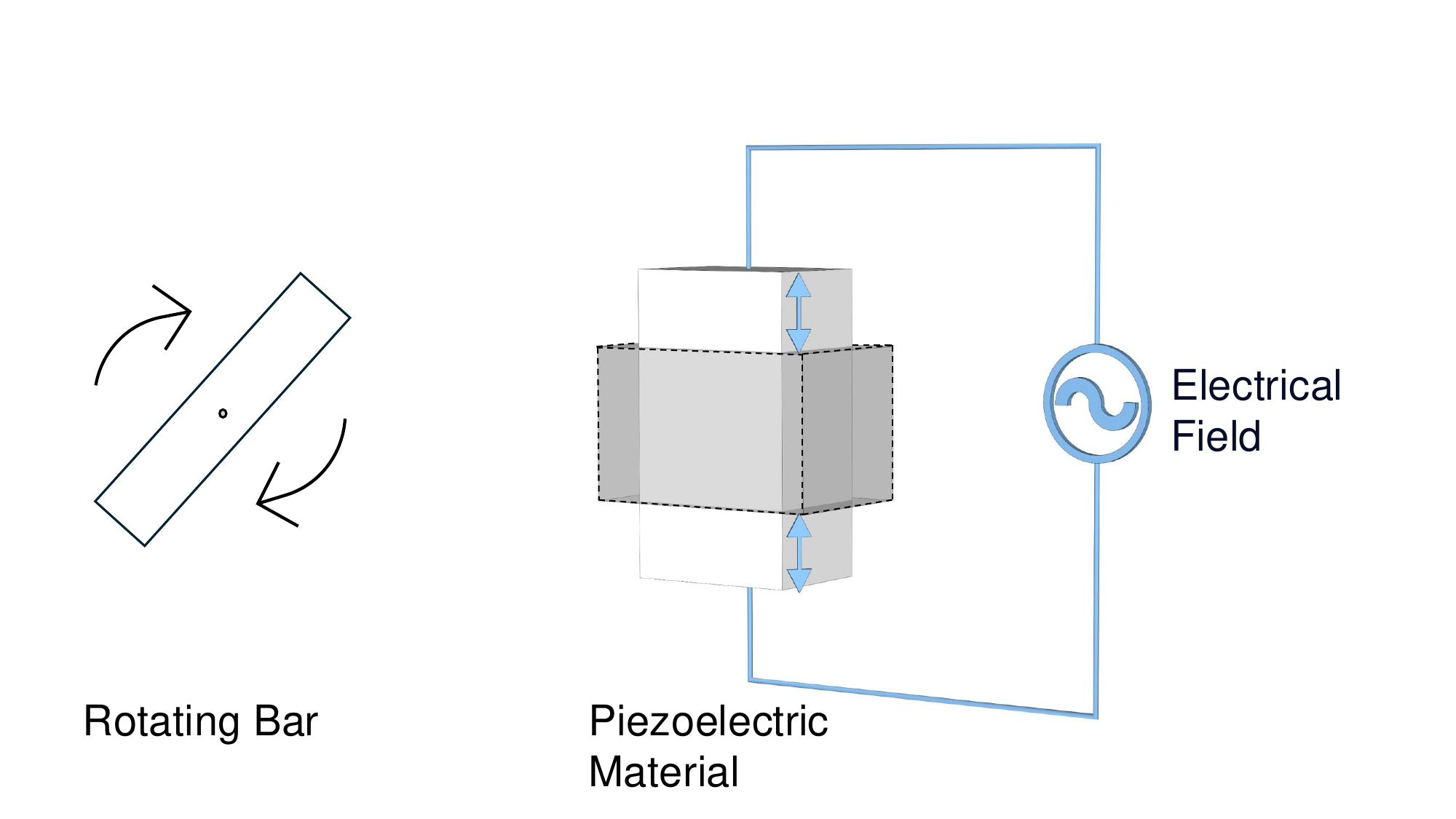}
    \caption{Two mechanical methods proposed in \cite{weber1960} for gravitational wave generation. (Left): The rotating rod method, generating low-frequency gravitational waves through asymmetric mass rotation.(Right): The piezoelectric crystal method, utilizing periodic mechanical stresses induced by an alternating electric field for higher frequency gravitational wave generation, offering greater efficiency.}
    \label{fig:mechnical}
\end{figure}

In 1960, the researcher in \cite{weber1960} proposed two mechanical methods for gravitational wave generation: the rotating bar method and the piezoelectric crystal method, as shown in Fig.~\ref{fig:mechnical}. The rotating bar method generates gravitational waves by rotating an asymmetric mass distribution at high speeds to excite changes in the mass quadrupole moment of the system. The fundamental equation describing the radiated power of a spinning rod is given by:
\begin{equation}
    P_g = -1.73 \times 10^{-59} I_m^2 \omega^6 \quad \text{(ergs/sec)},
\end{equation}
where $I_m$ is the moment of inertia and $\omega$ is the angular frequency. This equation indicates that the radiated power is extremely small, on the order of $10^{-59}$ ergs/sec, due to the weak coupling of gravitational waves to matter. The produced gravitational waves are also of low frequency, typically below 1 Hz, corresponding to wavelengths at least $10^6$ times the rod's length. Moreover, the quadrupolar nature of gravitational wave radiation further constrains this method, as gravitational waves arise only from the time-varying mass quadrupole moment ($\ddot{Q}$ $\neq$ 0) of a system. This means they cannot be emitted through monopolar (total mass changes) or dipolar (mass center movement) effects but require rapid, asymmetric mass changes, such as those in a rotating rod with uneven mass distribution. Combined with practical constraints like the material strength of the rod, which limits achievable rotation speeds, it becomes impossible to reach the high frequencies needed for significant gravitational wave emission, rendering gravitational wave generation via mechanical rotation inefficient and impractical for detection.

\subsubsection{Piezoelectric Crystal}In order to improve the efficiency, the piezoelectric crystal method was also proposed in \cite{weber1960}, in which an alternating electric field is applied to a piezoelectric crystal and the converse piezoelectric effect is utilized to trigger periodic mechanical stresses inside the crystal to generate gravitational waves. Compared with rotating rods, piezoelectric crystals can generate high-frequency gravitational waves driven by a high-frequency electric field. Although the generation rate is still limited, the increased efficiency of the piezoelectric crystal method makes it a more promising approach for subsequent gravitational wave generation studies.

\subsubsection{Mechanical Limitations}Further exploring mechanical methods, the efficiency of gravitational wave generation was analyzed in \cite{kraus1991}. It also introduced a gravitational wave antenna concept based on rotating bar. The research demonstrated that due to the quadrupole nature of gravitational waves, their radiation efficiency is significantly lower than that of electromagnetic waves, which exhibit a dipole structure. His calculations showed that even using a large rotating structure at high speeds would result in a power output far too weak to support effective communication. For example, consider a 500-tonne (\(5 \times 10^5 \, \mathrm{kg}\)) steel beam, 20 meters long, mounted on an axle to spin end-over-end at a rotational speed of 270 rpm—just below the speed at which centrifugal forces would cause the bar to break. The gravitational wave power output from such a structure, despite its enormous size and high rotational speed, is calculated to be only $10^{-29}$ Watts—much less than a picopicowatt. This insight highlighted the limitations of traditional mechanical methods and underscored the need for new approaches to overcome efficiency bottlenecks.

\subsubsection{Superfluid Sagnac}
\begin{figure}[!t]
    \centering
    \includegraphics[width=\linewidth]{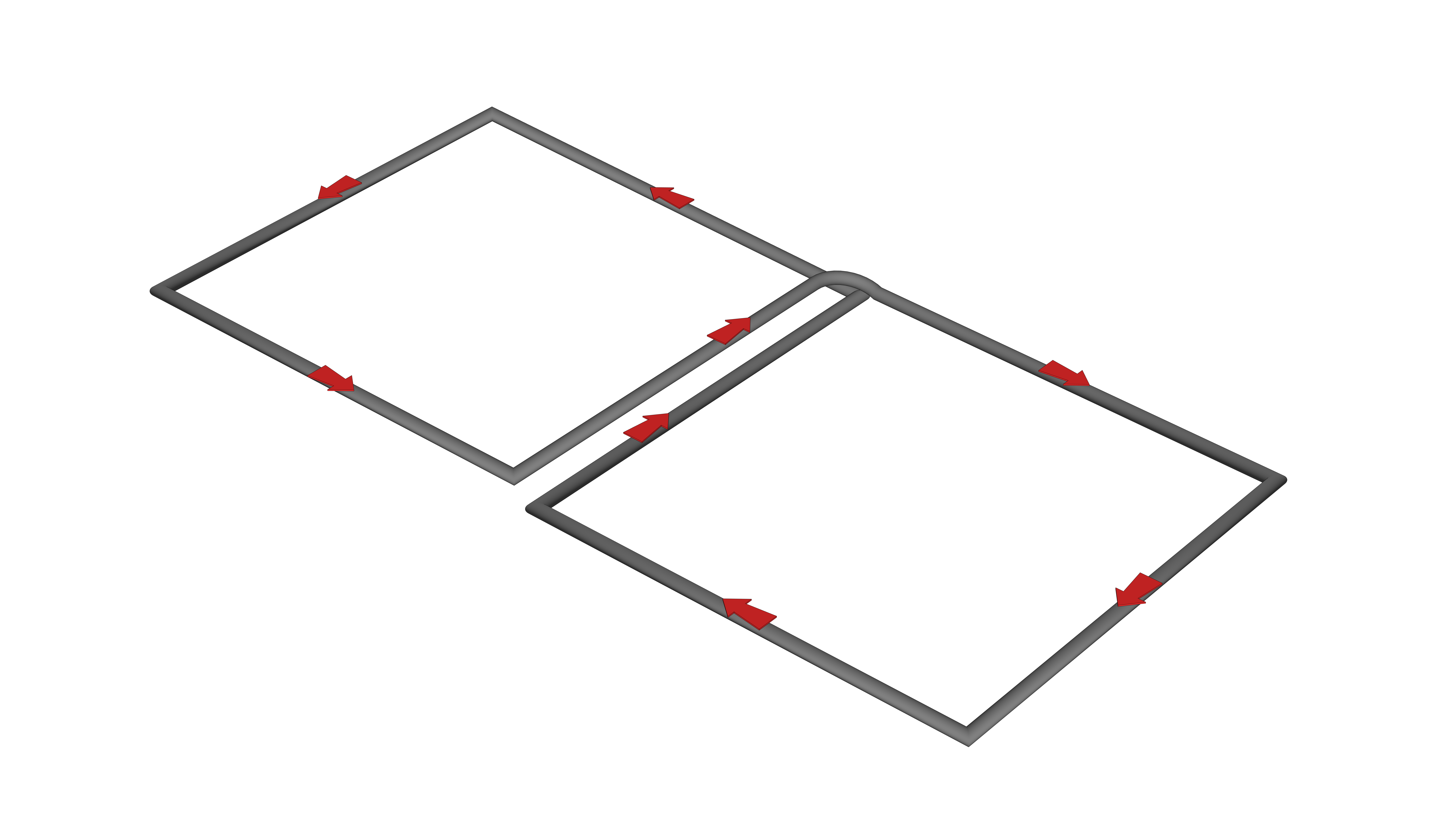}
    \caption{A schematic representation of the ``figure-eight" superfluid gravitational antenna, designed to generate gravitational waves using the Sagnac effect \cite{anandan1982}. Superfluid helium enable phase shifts within the rotating frame, where the Sagnac effect creates the necessary conditions for gravitational wave generation.}
    \label{fig:8}
\end{figure}

Another mechanical approach involved the use of superfluids and the Sagnac effect. In \cite{anandan1982}, the authors introduced the concept of a superfluid gravitational antenna. They proposed generating gravitational waves by creating phase shifts in a rotating superfluid-filled tube, as illustrated in ~Fig.~\ref{fig:8}. Although this method offered a novel experimental approach, the resulting gravitational wave amplitude was minuscule, limiting its practical applicability. The challenges stemmed from the difficulty in achieving sufficient mass movement and the precision required in manipulating superfluids to produce detectable gravitational waves.

\subsection{Superconducting and Particle Beam Collision Methods}

\subsubsection{Superconducting Resonance}Beyond mechanical methods, advancements in material science led researchers to explore superconducting materials for gravitational wave generation. The patent \cite{petlan2001} described a communication system that generates gravitational waves through resonance effects in superconducting materials, and the illustration is put in Fig.~\ref{fig:2sc}. This system utilizes two identical superconductors, where the density of one is periodically modulated to produce gravitational wave pulses, which are then detected by the second superconducting material. The applicant suggested using a specific alloy of rhodium and iridium to enhance resonance efficiency. Although this concept provided an innovative approach, it also faced practical challenges, including the need for specialized materials and extreme temperature control to maintain superconductivity.

\subsubsection{Quantum Transducer}Building on the properties of superconductors, \cite{chiao2007} proposed a method for using superconductors as quantum transducers to convert electromagnetic waves into gravitational waves, leveraging the ``Millikan oil drop" effect. This concept aimed to exploit the low impedance properties of superconductors for efficient signal conversion, enabling a bidirectional device that could generate and receive gravitational wave signals. This theory laid the foundation for subsequent experimental design and technical implementation. A recent work \cite{chiao2019} further developed this concept by designing a gravitational radiation communication system that employs a superconducting movable membrane between cylindrical superconducting cavities to achieve gravitational wave generation and detection through parametric amplification. However, the realization of this concept was hindered by material limitations and technical constraints inherent in working with quantum transducers and superconducting materials.

\subsubsection{Particle Beam Collisions}Another technique explored in this period involves particle beam collisions. In \cite{baker2004}, the author proposed a high-frequency gravitational wave (HFGW) generator that utilized particle beam collisions to produce gravitational waves, employing the concept of ``superradiance" to control frequency and direction of the emitted gravitational waves. Despite its innovative nature, this approach is limited by excessive energy requirements and insufficient theoretical validation.

\subsection{High-Power Lasers and EM Field-Based Approaches}

\subsubsection{High-Power Laser Radiation}
\begin{figure}[!t]
    \centering
    \includegraphics[width=\linewidth]{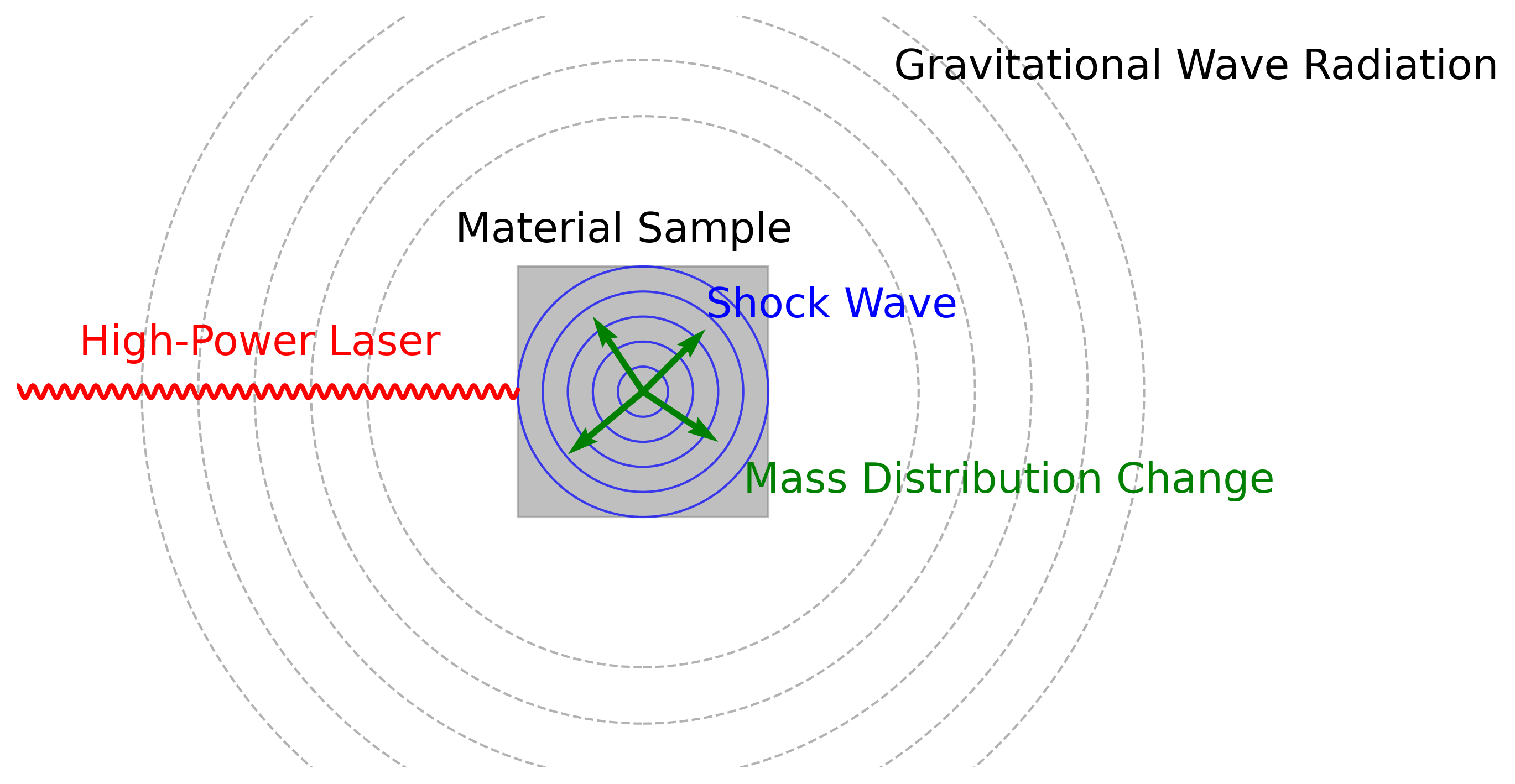}
    \caption{Illustration of high-power laser irradiation inducing shock waves in matter, leading to rapid changes in density and pressure, resulting in variations in mass distribution and the emission of gravitational wave radiation.}
    \label{fig:laser}
\end{figure}

Recent approaches aim to achieve higher-frequency and stronger-amplitude gravitational waves, one promising method involves using high-power laser radiation to perturb matter, thereby generating high-frequency gravitational waves (HFGWs). The research \cite{pustovoit2021} proposed that by irradiating matter with high-power lasers, the resulting shock waves cause dramatic changes in the density and pressure within the material, forming rapid variations in mass distribution, thereby producing gravitational wave radiation is shown in Fig.~\ref{fig:laser}. Here, the gravitational effect of the electromagnetic field itself can be neglected, and the response of the material to the laser pulses is considered the main source of gravitational waves. According to the quadrupole radiation formula in general relativity, changes in the quadrupole moment generate gravitational waves, and the intensity of these gravitational waves is proportional to the second time derivatives of the quadrupole moment of the source \cite{misner1973}:

\begin{figure}[!t]
    \centering
    \includegraphics[width=\linewidth]{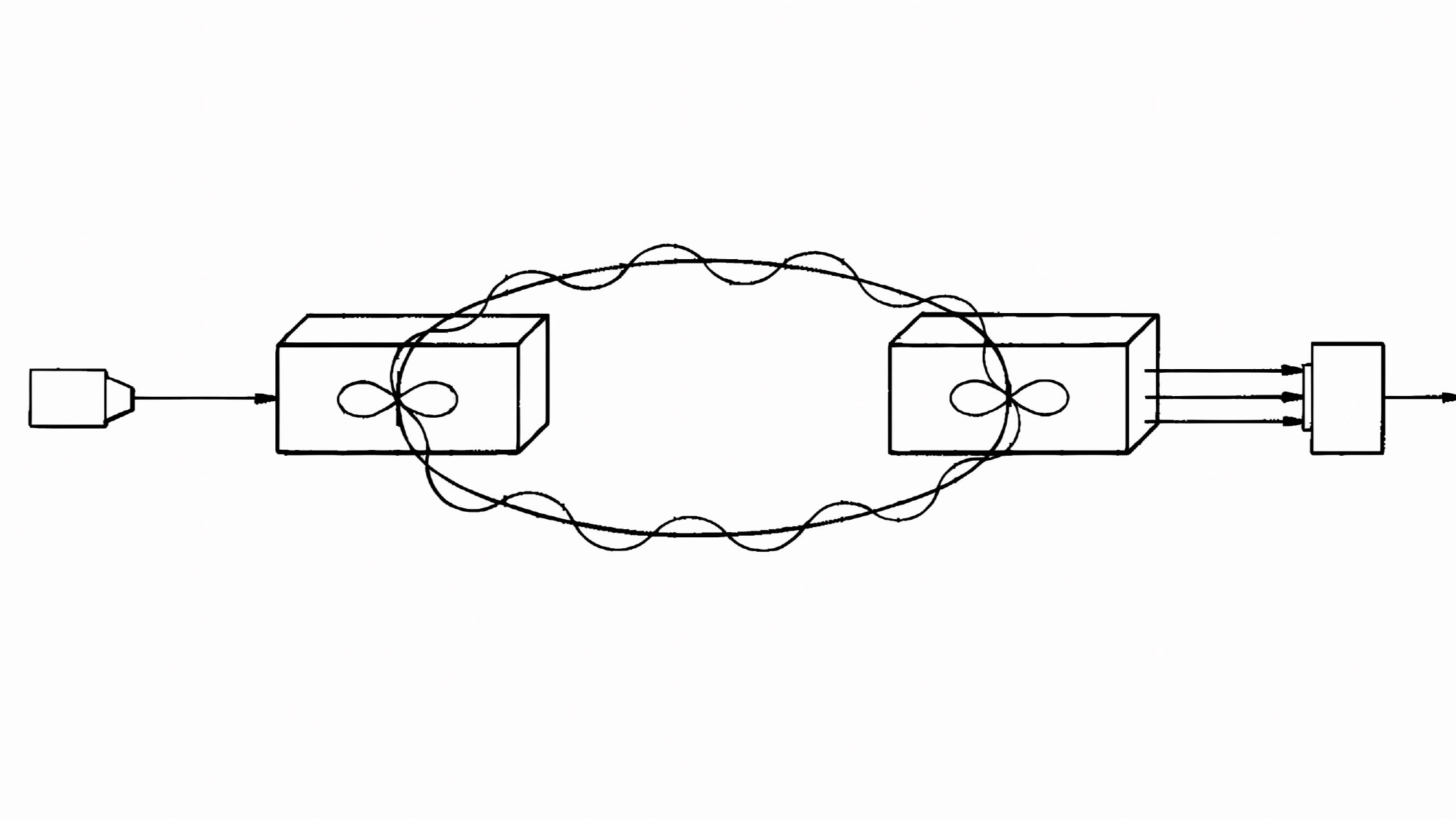}
    \caption{Gravitational wave communication system from \cite{petlan2001}, using two superconductors where density modulation in one generates gravitational waves detected by the other. The design leverages a rhodium-iridium alloy for efficiency but requires extreme temperature control for superconductivity.}
    \label{fig:2sc}
\end{figure}

\begin{equation}
    h^{\mu \nu} = \frac{2 G}{r c^4} \ddot{Q}^{\mu \nu}_{TT} \left( t - \frac{r}{c} \right)
    \label{eq:attenuation}
\end{equation} where \(h^{\mu \nu}\) is the perturbation of the metric tensor, or the amplitude of gravitational wave, \(G\) is the gravitational constant, \(r\) is the distance from the source to the observation point, \(c\) is the speed of light, \(\ddot{Q}^{\mu \nu}_{TT}\) is the second time derivative of the quadrupole moment in the transverse-traceless (TT) gauge, and \(t - \frac{r}{c}\) represents the retarded time. The quadrupole moment itself can be expressed as:
\begin{equation}
    Q^{\mu \nu}_{TT} = \int d^3 x \, \rho \left( x^\mu x^\nu - \frac{1}{3} \delta^{\mu \nu} r^2 \right)
\end{equation} where \(\rho\) is the mass density, \(x^\mu\) and \(x^\nu\) are the spatial coordinates, \(\delta^{\mu \nu}\) is the Kronecker delta ensuring tracelessness, and \(r^2 = x^\alpha x_\alpha\) is the squared spatial coordinate sum.

This proportional relationship holds as long as the wavelength of the emitted gravitational waves is significantly larger than the characteristic size of the source, denoted by \(R\).

Theoretical calculations show that the gravitational waves produced by this approach exhibit an amplitude in the range of \( h_0 \sim 10^{-40} \) \cite{pustovoit2021}, \cite{kadlecova2017}. The frequency of these waves depends on the properties of the laser and the shock waves, and while high frequencies can be achieved, the exact range is typically in the GHz domain. Due to the extremely small amplitude, the existing gravitational wave detectors cannot detect these signals. This indicates that this method faces significant challenges in practical applications, requiring a substantial increase in laser power or improvements in the sensitivity of detection technologies. 

\subsubsection{Electromagnetic Field Resonators}Additionally, approaches that utilize electromagnetic fields as sources for generating gravitational waves have garnered significant attention. In \cite{morozov2021}, researchers investigate the potential of producing high-frequency gravitational waves through standing electromagnetic waves confined within a Fabry-Perot resonator. For the optical frequency range \( w \sim 10^{14}\) to  \(10^{15}\) Hz, using realistic electric field strengths \( E_0 \sim 10^{6}\) V/m and resonator lengths $L$, the calculated gravitational wave amplitude \(h_0\) is approximately  \(10^{-46}\) to  \(10^{-44}\). 

Although the amplitude is extremely weak, this method provides a theoretical framework for generating gravitational waves using electromagnetic fields under laboratory conditions. Researchers point out that enhancing the strength of the electromagnetic field or optimizing the design of the resonator may improve the amplitude of the gravitational waves to some extent. However, due to material and technical limitations, there are practical difficulties in increasing the strength of the electromagnetic field.

\begin{table*}[ht]
\centering
\caption{Comparison of Gravitational Wave Communication Patents and Generation Methods}
\label{table_gw}
\begin{tabular}{|p{2cm}|p{1.3cm}|p{1cm}|p{5cm}|p{2cm}|p{4cm}|}
\hline
\textbf{Generation Approach} & \textbf{Inventor(s)} & \textbf{Publish Date} & \textbf{Technical Details} & \textbf{Application} & \textbf{Challenges} \\ \hline
\textit{Rotating Rod and Piezoelectric Crystal Methods} & J. Weber & 1960 & Rotating rod/bar: power outputs $\sim 10^{-59}$ ergs/sec; Frequencies range below 1 Hz. Piezoelectric crystal: high-frequency mechanical stresses for higher-frequency gravitational waves. \cite{weber1960}. & Experimental gravitational wave generation & Weak coupling to matter; low power output; material constraints. \\ \hline

\textit{Superfluid Gravitational Antenna} & J. Anandan & 1982 & Generates gravitational waves via phase shifts in a rotating superfluid-filled tube (Sagnac effect). Minuscule amplitude limits practical applicability \cite{anandan1982}. & Experimental gravitational wave generation & Precision required for manipulating superfluids; insufficient mass movement for detectable waves. \\ \hline

\textit{Gravitational Wave Antenna concept} & J. Kraus & 1991 & Large rotating structures (e.g., 500-tonne steel beam at 270 rpm) produce gravitational wave power of only $10^{-29}$ Watts \cite{kraus1991}. & Experimental gravitational wave generation & Inefficient radiation due to quadrupole structure; extremely low power output. \\ \hline

\textit{Superconducting blocks density modulation} & J. Petlan & Oct 9, 2001 & Generates and detects gravitational wave pulses through density modulation in superconducting blocks \cite{petlan2001}. & Long-distance wireless communication & Precision in density modulation; low-intensity waves; high noise levels. \\ \hline

\textit{Submicroscopic Energizable Elements} & R. Baker & Aug 31, 2004 & Generates gravitational waves by inducing ``jerk" motion in submicroscopic energizable elements \cite{baker2004}. & Communication, propulsion, physical theory testing & Precision control of jerk motion; low-energy gravitational waves. \\ \hline

\emph{High Power Laser Disturbance} & H. Kadlecov & Apr 18, 2017 & Gravitational waves generated by shock waves in matter induced by high-power lasers. Frequency range: $\sim 10^{14}$ to $10^{15}$ Hz. Amplitude range: $\sim 10^{-40}$ \cite{kadlecova2017}. & Experimental gravitational wave generation & Low amplitude makes detection impractical without significant improvements in laser power or detection technologies. \\ \hline

\textit{Superconducting Movable Membrane} & R. Chiao & May 28, 2019 & Amplifies gravitational wave signals using superconducting cavities with a movable membrane \cite{chiao2019}. & Gravitational wave communication systems & Maintaining superconducting materials at low temperatures; high-precision control of the membrane. \\ \hline

\emph{Standing Electromagnetic Waves} & A. Morozov & Dec 6, 2020 & Gravitational waves produced using standing electromagnetic waves in a Fabry-Perot resonator. Frequency range: $\sim 10^{14}$ to $10^{15}$ Hz. Amplitude range: $\sim 10^{-46}$ to $10^{-44}$ \cite{morozov2021}. & Experimental gravitational wave generation &  Weak amplitude and technical limitations in enhancing field strength and resonator design. \\ \hline
\emph{High-Power Twisted Light Beams} & E. Atonga & Aug 9, 2024 & Gravitational waves generated by high-power laser pulses carrying orbital angular momentum. Frequency range: $\sim 10^{16}$ Hz. Amplitude range: $\sim 10^{-37}$ to $10^{-36}$ \cite{atonga2024}. & Experimental gravitational wave generation & Generated waves remain weak and are beyond the sensitivity of current detection technologies. \\ \hline
\end{tabular}
\end{table*}

\subsubsection{Twisted Light Beam
}Another method using high-power twisted light beams (such as Bessel beams carrying orbital angular momentum) to generate gravitational waves has also attracted researchers' interest. In \cite{atonga2024}, researchers attempted to improve the radiation efficiency of gravitational waves by utilizing the special spatial structure and orbital angular momentum characteristics of the light beams. By adopting a reflective axicon optical system, high intensity twisted light beams are produced. Through fine-tuning the beam parameters, it is possible to manipulate the frequency, polarization, and direction of propagation of the generated gravitational waves.  

According to theoretical models, the gravitational waves generated by intense twisted light beams have a frequency around  \( \omega \sim 10^{16} \, \text{Hz} \). The gravitational wave amplitude is determined based on specific experimental parameters outlined in the study. For the scenario where the orbital angular momentum parameter \( l \) = 0, the peak strain amplitude of gravitational waves at a distance of \( 1.369 \, \mu\mathrm{m} \) from the source is estimated to be  \( h_0 \approx 1.90 \times 10^{-36} \)  when using a 1 PW laser pulse. In contrast, for  \( l \) = 1, the strain amplitude is calculated to be   \( h_0 \approx 4.69 \times 10^{-37} \) under the same conditions \cite{atonga2024}. These values reflect the extreme weakness of the gravitational waves generated, even with high-power laser pulses, making them challenging to detect with current gravitational wave detection technologies. These values illustrate the extreme weakness of the generated gravitational waves, making them difficult to detect with current technologies.

\subsection{Open Research Challenges}

To better illustrate the differences, comparison of these gravitational wave communication methods is included in Table~\ref{table_gw}. Each method has unique characteristics in terms of generation mechanisms, technical requirements, and inherent challenges.

\subsubsection{Mechanical Resonance and Rotational Techniques} Me-chanical methods for generating gravitational waves, such as rotating rods or bars with asymmetric mass distributions, face significant challenges due to extremely low radiation efficiency. The quadrupole nature of gravitational wave radiation requires time-varying mass quadrupole moments, but mechanical systems cannot achieve the necessary mass acceleration and frequency without exceeding material strength limits. Material limitations prevent achieving the high rotational speeds needed, as the centrifugal forces at such speeds would cause the structures to break apart. Additionally, the gravitational waves produced by these methods are of low frequency and exceedingly weak amplitude, making detection impractical with current technology.

\subsubsection{Superconducting and Particle Beam Collision Methods}
Methods involving superconductors and particle beam collisions encounter challenges related to material limitations and technical constraints. Superconducting techniques require maintaining extreme conditions, such as ultra-low temperatures, to preserve superconductivity, which increases technical complexity and cost. Manipulating superconducting materials and quantum transducers involves complex quantum effects that are difficult to control and utilize effectively for gravitational wave generation. Particle beam collision methods demand excessive energy inputs, yet the output gravitational wave energy remains minuscule, resulting in extremely low conversion efficiency. Furthermore, some of these concepts lack sufficient theoretical validation, necessitating further research to assess their feasibility.

\subsubsection{High-Power Lasers and EM Field-Based Approaches}
Approaches using high-power lasers and electromagnetic fields to generate gravitational waves aim to achieve higher frequencies and stronger amplitudes. However, the gravitational waves produced by these methods have extremely weak amplitudes (e.g., \( h_0 \sim 10^{-40} \)), far below the detection thresholds of existing detectors. Enhancing the amplitude is limited by practical constraints, such as the maximum achievable energy input and material limitations. Increasing laser power or optimizing the design of resonators may improve the amplitude to some extent, but significant challenges remain due to technical limitations and the extreme weakness of the generated signals. Current gravitational wave detectors cannot detect these high-frequency, low-amplitude signals, indicating a need for advancements in both generation and detection technologies.

\subsubsection{Frequency and Amplitude Limitations of Laboratory-Generated Gravitational Waves}
Laboratory methods for generating gravitational waves typically produce signals in the high-frequency range, approximately \( \omega \sim 10^{14} \) to \( 10^{16} \) Hz. These frequencies are significantly higher than those detectable by current gravitational wave detectors like LIGO, which are optimized for signals from tens to thousands of Hertz. Additionally, the amplitudes of gravitational waves produced by these laboratory methods are extremely weak, often in the range of \( h_0 \sim 10^{-46} \) to \( 10^{-44} \), making them undetectable with current technology. Furthermore, the amplitude of detected gravitational waves (\( h \)) diminishes with the range between the transmitter and receiver, following the linear inverse law \( h \propto 1/r \), as shown in~\eqref{eq:attenuation}. This attenuation over distance further complicates the detectability of such waves in practical scenarios. Developing new high frequency gravitational wave detection technologies or increasing the amplitude of generated gravitational waves is essential to bridge this gap. Considering the significant challenges in generating gravitational waves with both detectable amplitudes and frequencies compatible with existing detectors, it becomes crucial to explore advancements in detection technologies. Although current detectors are unable to capture the extremely weak and high-frequency gravitational waves produced in laboratory settings, understanding recent developments in gravitational wave detection methods is essential. The following section delves into both direct and indirect detection approaches, analyzing their contributions, limitations, and the potential they hold for future breakthroughs in gravitational wave communication.

\section{Gravitational Wave Detection}

\begin{figure}[!t]
    \centering
    \includegraphics[width=\linewidth]{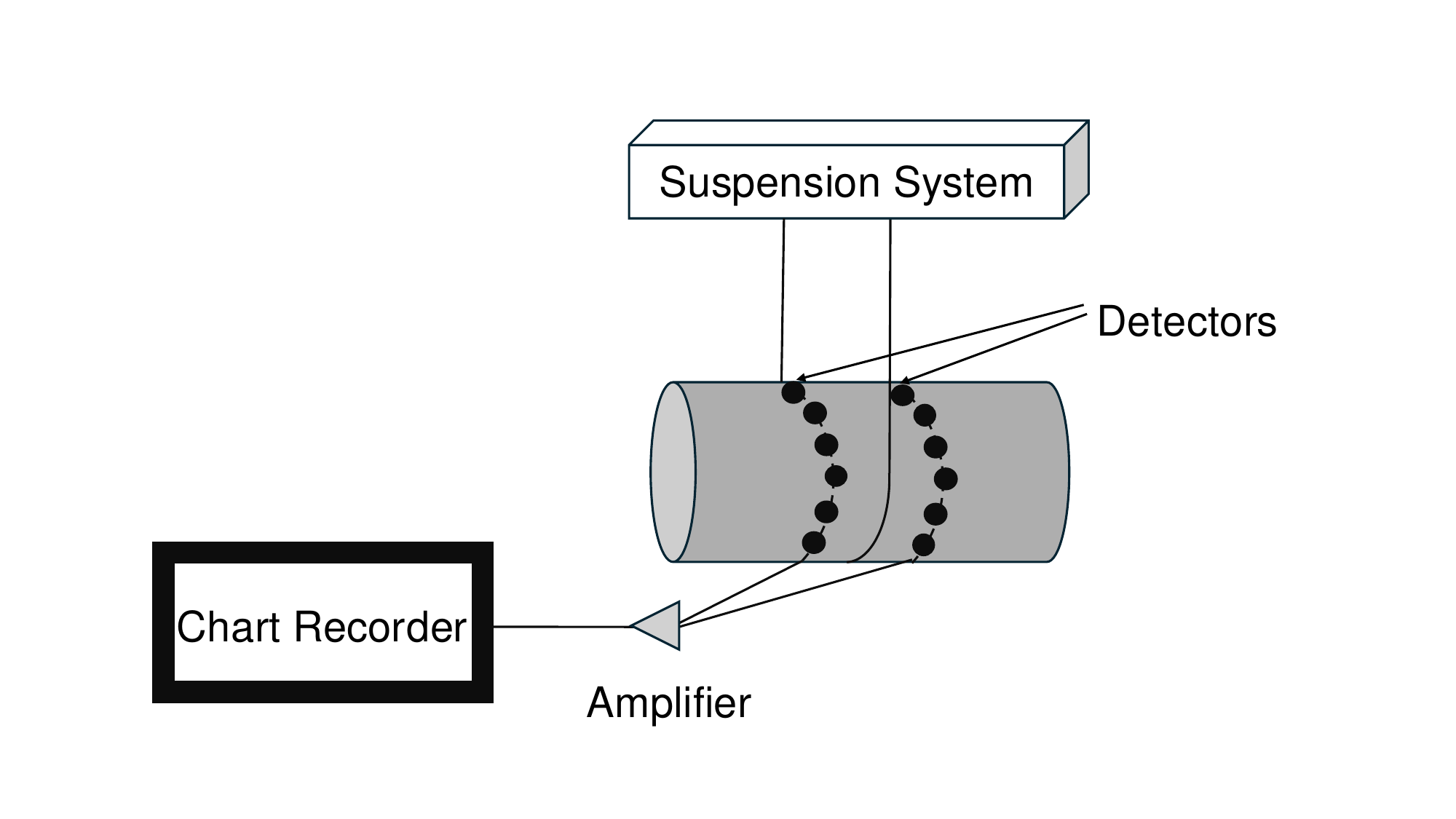}
    \caption{Diagram of Weber’s early gravitational wave detector. The detector, known as the Weber bar, senses resonant structural deformations caused by passing gravitational waves using piezoelectric crystal detectors attached along the bar, which convert strain into electrical signals recorded by a chart recorder. The entire detector system operates in a vacuum to reduce noise interference.}
    \label{fig:Weber bar}
\end{figure}

Gravitational wave detection is a pivotal aspect of gravitational wave communication, providing detailed analysis of these spacetime ripples. Since the 1960s, when Weber designed the first gravitational wave detector, known as the Weber bar as shown in Fig.~\ref{fig:Weber bar}, which aimed to detect gravitational waves through sensing resonant structural deformations, the field has seen significant advancements. Detection methods are broadly classified into direct and indirect approaches, each with unique techniques and challenges. Direct detection methods, particularly those based on laser interferometry, have become mainstream following the success of Laser Interferometer Gravitational-Wave Observatory (LIGO). Meanwhile, indirect methods explore gravitational wave interactions with electromagnetic fields and particles. To understand how gravitational wave detection has reached its current state, this section explores recent advancements in both direct and indirect gravitational wave detection methods, highlighting key technologies and unresolved challenges. 

\subsection{Laser Interferometry approach}

\subsubsection{Laser Interferometry}As the primary technique for direct gravitational wave detection, laser interferometry has played a transformative role in advancing our ability to observe these spacetime ripples. Since \cite{thorne1987} proposed using laser interferometers to measure minute spacetime distortions in 1987 , this technology has become the theoretical foundation for large-scale gravitational wave detectors. In 1991, \cite{kraus1991} highlighted the fundamental challenges faced by gravitational wave antennas in terms of energy conversion efficiency, which underscored the need for high-precision interferometry.

\begin{figure}[!t]
    \centering
    \includegraphics[width=\linewidth]{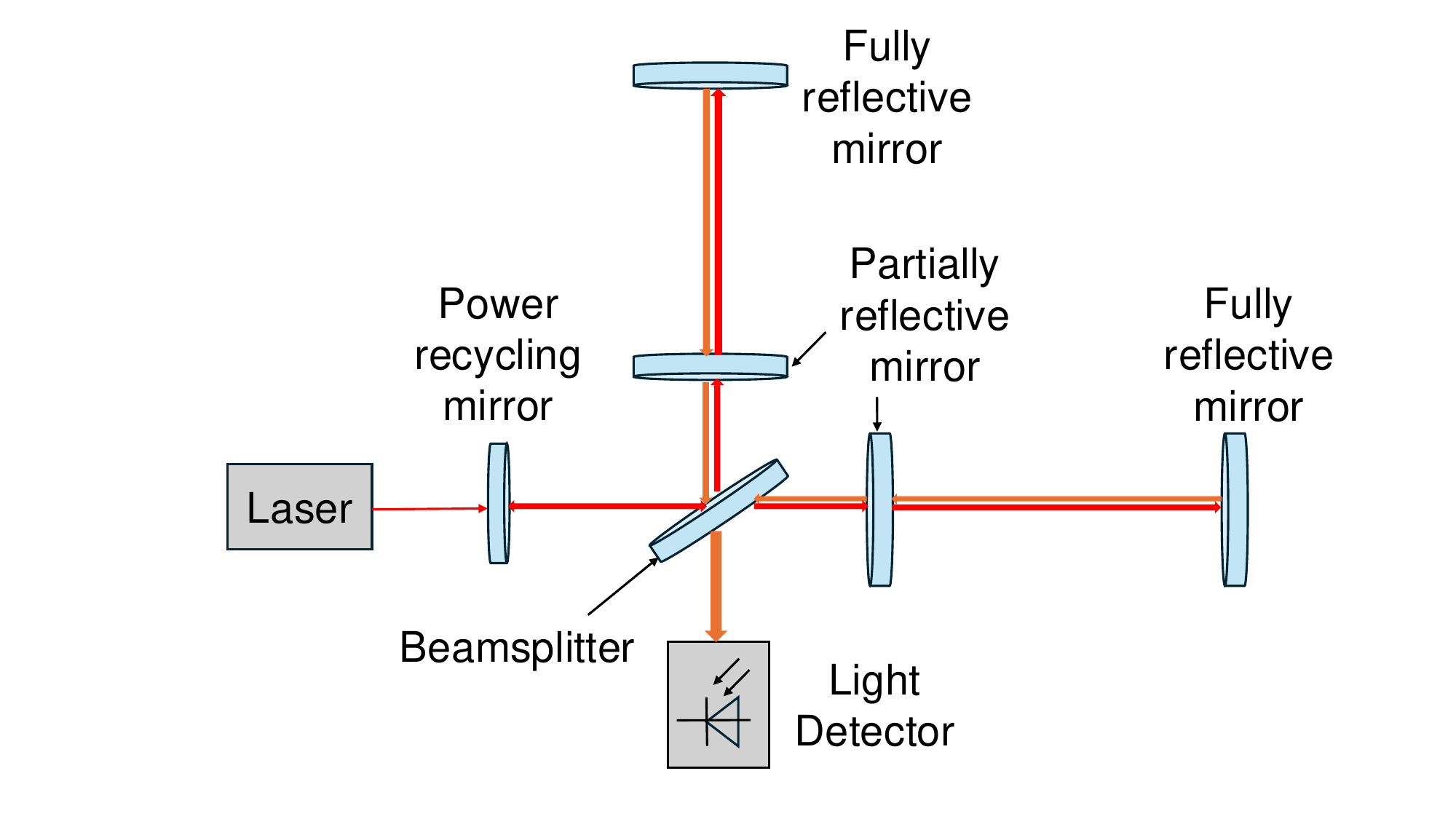}
    \caption{Simplified schematic of the LIGO interferometer, which is based on a high-precision Michelson interferometer \cite{giacomo1987}. Central to this technique, the interferometer splits laser light into two perpendicular arms using a beamsplitter. The light travels to fully reflective mirrors at the ends of each arm and returns to the beamsplitter. Gravitational waves cause minuscule changes in the arm lengths, creating interference patterns that the light detector captures, enabling the detection of these spacetime ripples.}
    \label{fig:LIGO}
\end{figure}

This approach, later pioneered by observatories like the LIGO and Virgo, detects gravitational waves in the range of tens to thousands of Hertz, with strain amplitudes down to \(10^{-21}\) \cite{abbott2016}, \cite{bond2016}. Central to this technique is the high-precision Michelson interferometer, as shown in~Fig.~\ref{fig:LIGO}. This interferometer detects changes in arm lengths caused by gravitational waves. Innovations such as Fabry–Perot cavities, power recycling, and signal recycling have been instrumental in enhancing sensitivity, despite significant challenges from quantum noise and environmental interference. Multi-stage suspension systems provide critical vibration isolation, stabilizing the delicate measurements necessary for accurate detections. This breakthrough ushered in a new era in gravitational wave astronomy and laid the groundwork for the subsequent application of deep learning and artificial intelligence in gravitational wave signal processing.

\subsubsection{Space-Based Interferometry (LISA)}
\begin{figure*}[!t]
    \centering
    \includegraphics[width=\linewidth]{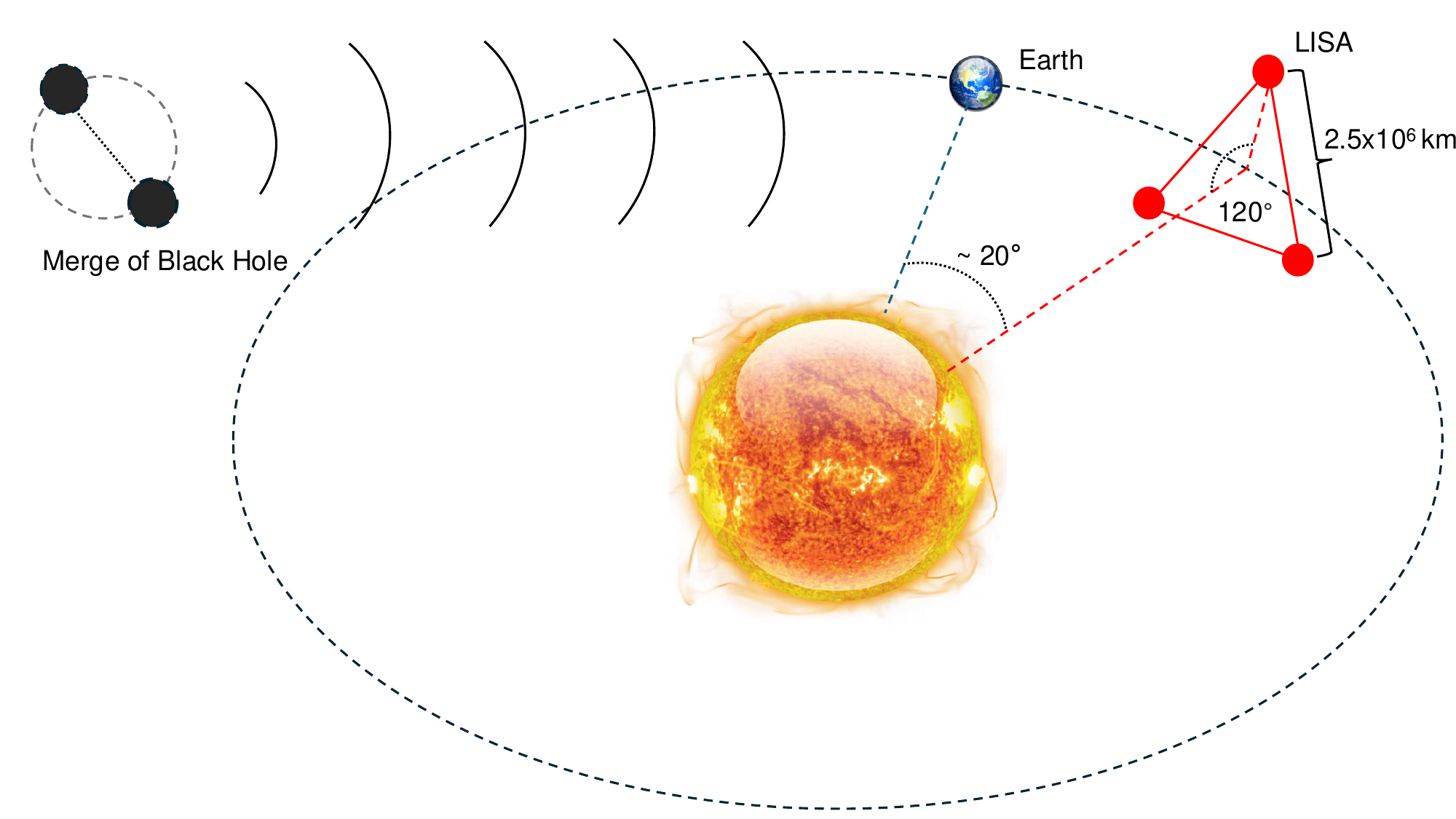}
    \caption{Schematics of LISA orbits. The constellation is trailing the Earth by about 20\textdegree, this orbital arrangement allows the constellation to be stable without further station keeping manoeuvres. The merge of black holes generates gravitational waves, which are subsequently detected by LISA within its sensitive frequency bands.}
    \label{fig:LISA}
\end{figure*}

Additionally, the European Space Agency's LISA project, proposed in \cite{danzmann2003}, suggests a space-based triangular interferometer array designed to detect low-frequency gravitational waves by combining laser interferometry with Doppler tracking. LISA aims to extend the capabilities of ground-based detectors by operating in the millihertz frequency range, which is inaccessible from Earth due to seismic and environmental noise. By placing three spacecraft in a triangular formation millions of kilometers apart, LISA will use laser interferometry to measure the minute changes in distance caused by passing gravitational waves, further advancing the field of gravitational wave detection. As illustrated in~Fig.~\ref{fig:LISA}, LISA's configuration allows for the detection of gravitational waves that cannot be observed by ground-based detectors.

\subsubsection{Dynamic Optical Spring Tracking}Building on these advancements, dynamic optical spring tracking has emerged as a key innovation for enhancing the sensitivity of quantum-limited interferometers \cite{aronson2024}. By leveraging the optical spring effect, this technique dynamically tunes the resonant frequency of a detuned optical cavity to match the evolving frequency of gravitational wave signals, effectively minimizing quantum noise and surpassing the standard quantum limit (SQL) in critical ranges.

Experiments demonstrate substantial improvements in signal-to-noise ratio (SNR), with average gains of 8.5 and peaks up to 40 at specific frequencies. Unlike frequency-dependent squeezing, which is sensitive to optical losses, optical spring tracking remains robust in high-loss scenarios, offering an alternative for noise suppression in next-generation detectors. Its adaptability extends the bandwidth of ground-based observatories like LIGO and aligns with space-based platforms like LISA.

This approach addresses longstanding challenges in quantum noise suppression and signal tracking, significantly advancing the detection capabilities of laser interferometers and opening new possibilities for gravitational wave astronomy.

\subsection{Deep Learning for Direct Detection}

\subsubsection{MSNRnet-2 Framework}In recent years, deep learning has shown immense potential in gravitational wave signal processing, particularly in signal detection, noise reduction, and decoding. Although many studies initially targeted gravitational wave detection from astronomical events, these advancements have laid a robust foundation for applying deep learning techniques to gravitational wave communication.

Leveraging this progress, researchers have introduced an advanced deep learning framework called MSNRnet-2 for directly denoising and detecting gravitational wave signals \cite{ma2024}. Inspired by the original MSNRnet and generative adversarial networks (GANs), this upgraded method simultaneously trains denoising and discrimination networks. It aims to transform ``signal + noise" data into pure signal shapes while converting pure noise data into non-astrophysical signal shapes, thereby reducing the tendency to misclassify noise as signals. By testing MSNRnet-2 on the O3b observational data from LIGO's Hanford and Livingston interferometers, it significantly reduced errors in signal-to-noise ratio (SNR) predictions, enhanced denoising performance, increased the detection rate by 1.3\%, and produced no false alarms \cite{ma2024}. Compared to traditional matched filtering methods, this deep learning framework exhibits higher efficiency and accuracy in handling complex gravitational waveforms that include higher-order modes and orbital eccentricities, solving the computational limitations often encountered by matched filtering algorithms that rely on template banks. The successful application of MSNRnet-2 demonstrates the immense potential of deep learning techniques in gravitational wave reception and processing, offering new technological means for signal detection in third-generation gravitational wave detectors. 

\subsubsection{Automated machine learning}Moreover, the widespread application of artificial intelligence and deep learning technologies is comprehensively revolutionizing the field of gravitational wave detection. Researchers have applied deep learning models such as convolutional neural networks (CNNs), autoencoders, and recurrent neural networks (RNNs) to the detection and analysis of various types of gravitational waves \cite{prajapati2024}. By utilizing automated machine learning (AutoML) techniques, these models can adaptively optimize their architectures to improve detection efficiency and accuracy. Testing on simulated and real data from ground-based detectors like LIGO, Virgo, and KAGRA, these AI models have significantly enhanced detection accuracy and greatly reduced computation time when processing gravitational wave signals from sources such as binary black hole mergers, binary neutron star mergers, and black hole-neutron star mergers. Particularly in complex noisy environments, deep learning models can still accurately identify gravitational wave events. Although the authors point out that existing models involve trade-offs in aspects like detection, parameter estimation, denoising, and glitch reduction, statistical analyses show that AI models perform better than traditional matched filtering methods in terms of detection rate and error reduction.

\subsubsection{Multimodal Machine Learning in Multi-Messenger Astronomy}Additionally, the application of multimodal machine learning (MML) techniques has brought new breakthroughs to multi-messenger astronomy in the field of gravitational wave reception. Focusing on binary neutron star merger events, researchers have proposed an MML method that combines gravitational wave signals with gamma-ray burst (GRB) data \cite{cuoco2021}. This method processes time-frequency images of gravitational waves and gamma-ray light curves, utilizing CNNs to extract and fuse features, achieving more accurate predictions of astrophysical parameters such as redshift. Compared to traditional methods that analyse gravitational wave and electromagnetic signals separately, this innovative approach integrates multiple data modalities within a single machine learning pipeline, significantly improving the analysis efficiency and accuracy of multi-messenger events. Although current research is primarily based on simulated data, results indicate that MML has enormous potential in predicting key parameters of multi-messenger events \cite{cuoco2021}. In the future, by testing on real observational data and incorporating more diverse astrophysical sources (such as neutrinos), MML is expected to further enhance gravitational wave reception and analysis capabilities, expanding the application scope of AI technologies in astrophysical research. 
  
Regardless of the significant progress achieved by these deep learning and artificial intelligence techniques in the field of gravitational wave reception, they still face some important challenges. For instance, MSNRnet-2 exhibits suboptimal denoising performance when dealing with events of extremely low SNR (e.g., SNR between 5 and 6), often producing outputs that resemble noise shapes. This indicates a need for further optimization in handling low SNR strain data. Additionally, the complex noise environment poses severe challenges to AI models. Gravitational wave detection data is intermingled with various noise sources such as instrumental vibrations, thermal noise, and seismic activities, which can obscure the true gravitational wave signals, especially in cases of weak or non-stationary signals. The quality and diversity of training data also limit the generalization ability of the models. Due to the rarity of gravitational wave events, obtaining large amounts of high-quality labelled data is difficult, leading to potential poor performance when models encounter rare events or those significantly different from the training data. Noise and glitches in the training data can interfere with the learning effectiveness of the models. High computational costs and the need for real-time processing are additional obstacles that need to be overcome, particularly when integrating these complex AI models into existing gravitational wave observation systems. Regarding multimodal machine learning methods, practical applications may encounter challenges such as data complexity and noise interference. Therefore, to reliably apply these advanced technologies in actual observations, further research and model improvements are necessary to overcome the aforementioned challenges and enhance the detection accuracy and efficiency of gravitational wave signals. 

\subsection{Electromagnetic Interactions-Based Indirect Detection}

Indirect detection methods, which involve detecting interactions between gravitational waves and electromagnetic fields, magnons, or photons, offer feasible pathways for realizing gravitational wave communication. These methods not only overcome the sensitivity limitations of direct detection but also provide new technical means for information transmission and extraction. 


\subsubsection{Inverse Gertsenshtein Effect}This research \cite{li2013} proposes methods to detect HFGWs by exploring the electromagnetic responses induced in a background electromagnetic field. In their earlier work, they investigated the inverse Gertsenshtein effect \cite{gertsenshtein1962} for HFGW detection, targeting frequencies from GHz to THz with detectable amplitudes as low as \(10^{-21}\). This effect involves the conversion of gravitational waves into electromagnetic waves within a strong magnetic field. By coupling the inverse Gertsenshtein effect with coherent electromagnetic waves and employing superconducting microwave cavities and synchro-resonance systems, the detection sensitivity can be improved by one to two orders of magnitude under specific conditions. For example, in the coupling system with a coherent plane electromagnetic wave, the first-order perturbative electromagnetic signal flux can reach $5.0 \times 10^{12} \, \text{s}^{-1}$, which is approximately $10^{11}$ times larger than the second-order signal flux produced by the pure inverse Gertsenshtein effect. However, due to the presence of significant background photon noise, this improvement does not fully translate into a proportional enhancement in overall sensitivity. Instead, the detectable gravitational wave amplitude ($h_{\text{min}}$) improves to a level of $h_{\text{min}} = 1.8 \times 10^{-24}$ for gravitational waves with a frequency of $3 \times 10^{12} \, \text{Hz}$. This represents a substantial enhancement compared to the minimum detectable amplitude for gravitational waves under the pure inverse Gertsenshtein effect, which is limited to amplitudes larger than $10^{-21}$ under similar laboratory conditions. This research laid the groundwork for more sensitive detection technologies and highlighted the complementary role electromagnetic detection could play alongside existing detectors focused on lower frequencies. 

\subsubsection{Synchro-Resonance}Building upon this foundation, \cite{li2020} focuses on detecting gravitational waves by examining the Electromagnetic (EM) response induced by HFGWs in a background electromagnetic field. This study targets frequencies ranging from \( 10^{8}\) to  \(10^{12}\) Hz and explores the detection of gravitational wave amplitudes in the range of \( 10^{-27}\) to  \(10^{-21}\) and \( 10^{-36}\) to  \(10^{-31}\) depending on the source. Utilizing electrodynamics in curved spacetime, they derived analytical solutions for the perturbed EM fields resulting from the interaction between gravitational waves and background EM fields. Specifically, they designed a Three-Dimensional Synchro-Resonance System (3DSR) capable of distinguishing different polarization states of HFGWs. By detecting these EM signals, the system can provide detailed information about the characteristics and origins of the gravitational waves, thereby enhancing our understanding of these waves and their sources. 

\subsubsection{First-Order Electromagnetic Response}Lastly, \cite{zheng2022} proposes a method for detecting HFGWs by utilizing first-order electromagnetic responses generated in a modulated magnetic field. Unlike the popular approach based on the inverse Gertsenshtein effect \cite{palessandro2023}, \cite{he2024}, which relies on second-order electromagnetic responses and often results in weak signals, their method enhances detection sensitivity by focusing on first-order electromagnetic responses. They designed an experimental system consisting of a static high magnetic field modulated by a weak alternating magnetic field and utilized single-photon detectors to capture the electromagnetic signals induced by passing gravitational waves. By modulating the magnetic field, they were able to amplify the interaction between gravitational waves and electromagnetic fields, making it feasible to detect HFGWs in the MHz and GHz bands. This approach provides a potential method for the indirect detection of gravitational waves and offers insights into verifying cosmological models through electromagnetic signals. 

\subsection{Indirect Detection Using Magnons}
Beyond electromagnetic interactions, other researchers have proposed novel methods for the indirect detection of HFGWs. \cite{ito2023} explores the use of magnons—collective spin excitations in ferromagnetic materials—as a means for detecting high-frequency gravitational waves. Their study demonstrates that magnon systems are particularly sensitive to high-frequency GWs in the gigahertz range, making them a promising approach for such detections. By refining theoretical models and expanding the application of Fermi normal coordinates, they improved the detection sensitivity to approximately \( h_c \sim 10^{-20} \). While primarily aimed at detecting GWs, the methodology and increased sensitivity achieved through the use of magnons, especially when combined with quantum sensing techniques, could pave the way for future explorations in high-frequency GW applications. This work highlights the potential of magnon systems as a complementary tool in the search for high-frequency GWs, possibly contributing to the broader field of GW research and detection strategies. 

\subsection{Optical Frequency Modulation Approach}

\begin{table*}[!t]
\renewcommand{\arraystretch}{1.3}
\caption{Comparison of Gravitational Wave Detection Methods}
\label{table_gw_detection}
\centering
\footnotesize
\begin{tabular}{|p{4cm}|p{3cm}|p{3cm}|p{6cm}|}
\hline
\textbf{Method} & \textbf{Frequency Range of \newline Detectable GW (Hz)} & \textbf{Amplitude Range ($h_0$) \newline of Detectable GW} & \textbf{Description of Detection Method} \\ \hline
\emph{Laser Interferometer Detection \newline (LIGO, Virgo, KAGRA)} & $\sim 10$ to $10^{3}$ & $\sim 10^{-21}$ & Detecting gravitational waves by monitoring small changes in space distance \cite{abbott2016}, \cite{abbott2017}.\\ \hline
\emph{Deep Learning Enhanced Detection} & $\sim 10$ to $10^{3}$ &  $\sim 10^{-21}$ & Improves detection accuracy in noisy environments \cite{ma2024}, \cite{prajapati2024}.\\ \hline
\emph{Indirect Detection via EM Responses} & $\sim 10^{8}$ to $10^{12}$ & $\sim 10^{-36}$ to $10^{-31}$ or  \newline  $\sim 10^{-27}$ to $10^{-21}$& Uses interactions between gravitational waves and electromagnetic fields \cite{li2013}, \cite{li2020}. \\ \hline
\emph{Magnon-Based Detection} & $\sim 10^{9}$ & $\sim 10^{-20}$ & Utilizes magnons for high-frequency gravitational wave detection \cite{ito2023}. \\ \hline
\emph{Optical Frequency Modulation Techniques} & $\sim 10^{6}$ to $10^{9}$ & Dependent on setup & Detects frequency modulation of photons induced by gravitational waves \cite{bringmann2023}. \\ \hline
\end{tabular}
\end{table*}

In addition to electromagnetic and magnon-based methods, optical frequency modulation offers innovative approaches for the indirect detection of HFGWs. \cite{bringmann2023} proposes detecting high-frequency gravitational waves through the frequency modulation of photons induced by gravitational waves. They utilized strong laser beams, optical frequency demodulation, and atomic clock technology to precisely measure the photon frequency shifts caused by gravitational waves. The proposed methods target gravitational waves in the MHz–GHz frequency range, where conventional detectors struggle to achieve sufficient sensitivity. The study highlights that the strain amplitude \(h_{+}\) of gravitational waves plays a critical role in the induced modulation effects, which can be detected by observing frequency shifts in the photon spectrum. These high-precision optical measurement methods enhance the sensitivity of gravitational wave detection across a broad frequency range, supporting future experimental systems. 

However, despite these promising advancements, these innovative methods face several common challenges. A primary obstacle is the extremely weak amplitude of HFGWs, which can range from \( 10^{-27}\) to  \(10^{-20}\) depending on the source, making it difficult to distinguish the gravitational wave signals from background electromagnetic noise. Achieving the necessary sensitivity often requires extremely high magnetic fields and advanced superconducting cavities with high-quality factors, which current technology struggles to provide. For instance, generating magnetic fields strong enough and maintaining the required conditions in superconducting cavities are significant technical hurdles. Additionally, precise control and alignment of experimental setups—such as maintaining stable and precise measurement systems in optical frequency modulation techniques or accurately aligning magnon detectors to the gravitational wave propagation direction—are challenging due to potential fluctuations and external interferences. Geometric constraints of detection cavities or materials can limit the effective frequency range, and strong background noise from zero-order electromagnetic responses or conventional electromagnetic induction can obscure the weak signals induced by HFGWs. While techniques like wave-impedance matching and advanced optical filtering can help mitigate these issues, they add complexity to the system implementation. These challenges highlight the need for further advancements in detection technology, noise reduction strategies, and experimental techniques to enhance the feasibility and reliability of indirect HFGW detection methods. 

\subsection{Open Research Challenges}
To facilitate comparison of the various gravitational wave detection methods, Table~\ref{table_gw_detection} summarizes their key characteristics, including frequency ranges and sensitivity levels. 

\subsubsection{Sensitivity and Technology Gaps}While direct detection methods improve accuracy in complex environments, they demand significant resources and require optimization for high-frequency gravitational waves. Indirect methods enhance sensitivity, but face challenges with low-frequency signals. High-frequency gravitational waves, often generated by smaller masses or scales, are feasible for artificial production under laboratory condition. But they remain undetectable due to their low amplitudes and the mismatch with current detector sensitivities. These limitations directly impact not only the study of gravitational waves but also their potential applications, such as in communication systems. Advancing detection technologies or aligning wave generation with detector capabilities is crucial to unlocking these possibilities.

\subsubsection{Frequency and Amplitude Impacts}Moreover, while the existing research primarily focuses on detecting gravitational waves of different amplitudes and frequencies—and emphasizes the importance of frequency and amplitude in detection sensitivity—it does not further explore their impact on gravitational wave communication itself. In traditional communication systems, the carrier frequency directly affects the available bandwidth, thereby influencing the data transmission rate; higher frequencies typically allow for larger bandwidths and higher data rates. Similarly, amplitude reflects the signal strength, affecting its detectability and resilience to noise. Can similar principles be applied to gravitational wave communication? How do gravitational wave frequency and amplitude affect the bandwidth, data rate, and signal propagation characteristics of gravitational wave communication?

\subsubsection{Key Parameters Definition}To answer the above questions, we need to first clarify the key parameters in gravitational wave communication:
\begin{itemize}

    \item \textbf{Bandwidth}: Refers to the range of frequencies over which gravitational waves can effectively carry information. Mathematically, it can be expressed as:
    \[
       B = f_{\text{high}} - f_{\text{low}}
    \]
    where \(f_{\text{high}}\) and \(f_{\text{low}}\) represent the upper and lower frequency limits. Higher frequencies might offer broader bandwidths, but the weak amplitudes and current limitations in detection technology constrain their practical utilization.
    
    \item \textbf{Data Rate}: Represents the amount of information transmitted per unit time. Higher carrier frequencies theoretically allow for higher data rates; however, challenges such as low signal amplitudes and noise interference during propagation significantly limit this potential. The data rate can be expressed mathematically as:
    \[
       R = B \cdot \log_2(1 + \text{SNR})
    \]
    A wider bandwidth and a higher SNR directly translate to a higher data rate, making these parameters essential for efficient gravitational wave communication.

    \item \textbf{Signal Propagation Characteristics}: Encompass how gravitational waves travel through space, including attenuation, waveform distortion, and polarization changes caused by interactions with cosmic structures like magnetic fields, interstellar media, or dense matter. These factors critically affect signal integrity and overall communication performance.
\end{itemize}

\subsubsection{Challenges in Sensitivity, Bandwidth, and Propagation Modeling}Above parameters are directly determined by the frequency and amplitude of gravitational waves. However, the relationship between these parameters and the practical challenges of gravitational wave communication remains unexplored. For instance, how do gravitational wave frequency and amplitude affect achievable bandwidth and data rate? How do propagation characteristics, influenced by cosmic structures, shape the feasibility of long-distance communication? Addressing these questions requires tackling several key technological and theoretical challenges:

\begin{itemize}
    \item \textbf{Detection Sensitivity for High-Frequency Gravitational Waves:} Current detection methods are limited in their ability to capture high-frequency, low-amplitude gravitational waves, which are critical for exploring their potential in communication systems. Existing detectors, designed primarily for astrophysical sources, lack the sensitivity required for artificially generated or weak signals. Research should focus on designing detectors capable of operating across broader frequency and amplitude ranges.
    
    \item \textbf{Quantifying Bandwidth and Optimizing Data Rate:} Gravitational wave communication lacks a clear framework for defining and quantifying bandwidth, particularly in relation to carrier frequency. Similarly, achieving high data rates faces significant hurdles due to the weak amplitudes and noise interference during signal propagation. These limitations necessitate the development of tailored modulation schemes to improve transmission efficiency.
    
    \item \textbf{Modeling Propagation Effects in Complex Environments:} The interaction of gravitational waves with cosmic structures, such as magnetic fields, dense matter, and interstellar media, introduces challenges including attenuation, phase distortion, and polarization shifts. These effects not only degrade signal quality but also complicate decoding, especially in interstellar or intergalactic scenarios where scalability and energy efficiency become critical. Additionally, unique noise sources, such as cosmic background radiation, thermal gravitational noise, and overlapping gravitational wave signals, exacerbate the difficulty of signal extraction. Developing comprehensive channel models is essential to ensure reliable and efficient detection in these environments.
\end{itemize}

\subsubsection{Detection \& Estimation Theory Under Extreme Low-SNR Conditions} Although existing gravitational wave detectors have achieved remarkable sensitivity for astrophysical events, transitioning from one-off event detection to robust gravitational wave communication requires advanced detection and estimation techniques capable of handling continuous or quasi-continuous low-amplitude signals. Traditional matched-filtering methods, largely tailored to short transient waveforms (e.g., black hole mergers), must be reimagined for iterative or adaptive detection schemes that can incorporate additional communication parameters. Moreover, non-stationary and non-Gaussian noise—ranging from seismic and thermal fluctuations to quantum shot noise—poses further challenges, necessitating sophisticated statistical modeling. This endeavor remains largely unexplored, forming a pivotal open research challenge that requires closer collaboration between gravitational wave physicists and signal processing experts.

By addressing these interconnected challenges, future research can build a solid foundation for practical gravitational wave communication systems, pointing the way toward the development of novel
generation and detection methods that can bridge the current technological gaps.

\section{Gravitational Wave Modulation}

Building on the significance of gravitational wave frequency and amplitude, effective modulation of these waves becomes critical for advancing gravitational wave communication. Gravitational wave modulation refers to the manipulation of wave properties such as amplitude, frequency, and phase to enhance compatibility with detectors and improve communication efficiency. These techniques address challenges in generating and detecting gravitational waves, offering novel approaches for transmitting information through spacetime. Recent studies have explored diverse methods, including astrophysical phenomena-based amplitude modulation (AM), dark matter-induced frequency modulation (FM), superconducting material manipulation, and nonmetricity-based theoretical approaches. The following subsections will delve into these methods, highlighting their principles, challenges, and future directions.

\subsection{Amplitude Modulation Based on Astrophysical Phenomena}

In \cite{qiu2022}, time-domain numerical simulations and the finite element method are employed to solve the wave equation in curved spacetime. This approach allows the investigation of the modulation effects of binary black hole lenses on gravitational wave signals. They propose a dynamic lens system, where time-evolving gravitational potentials modulate the gravitational wave signal. This mechanism resembles AM in radio communications and reveals unique features of wave propagation in complex lensing systems.

When gravitational waves pass through a binary black hole lens system from the source, the dynamic potential of the lens periodically changes the amplitude of the gravitational waves, resulting in sideband effects in the frequency domain. These sideband features appear on both sides of the main frequency, and their frequencies are generated by the convolution of the source gravitational wave frequency and the lens system frequency, expressed as \( f_{\text{sideband}} = f_{\text{source}} \pm 2 \times f_{\text{lens}} \). 

\begin{figure}[!t]
    \centering
    \includegraphics[width=\linewidth]{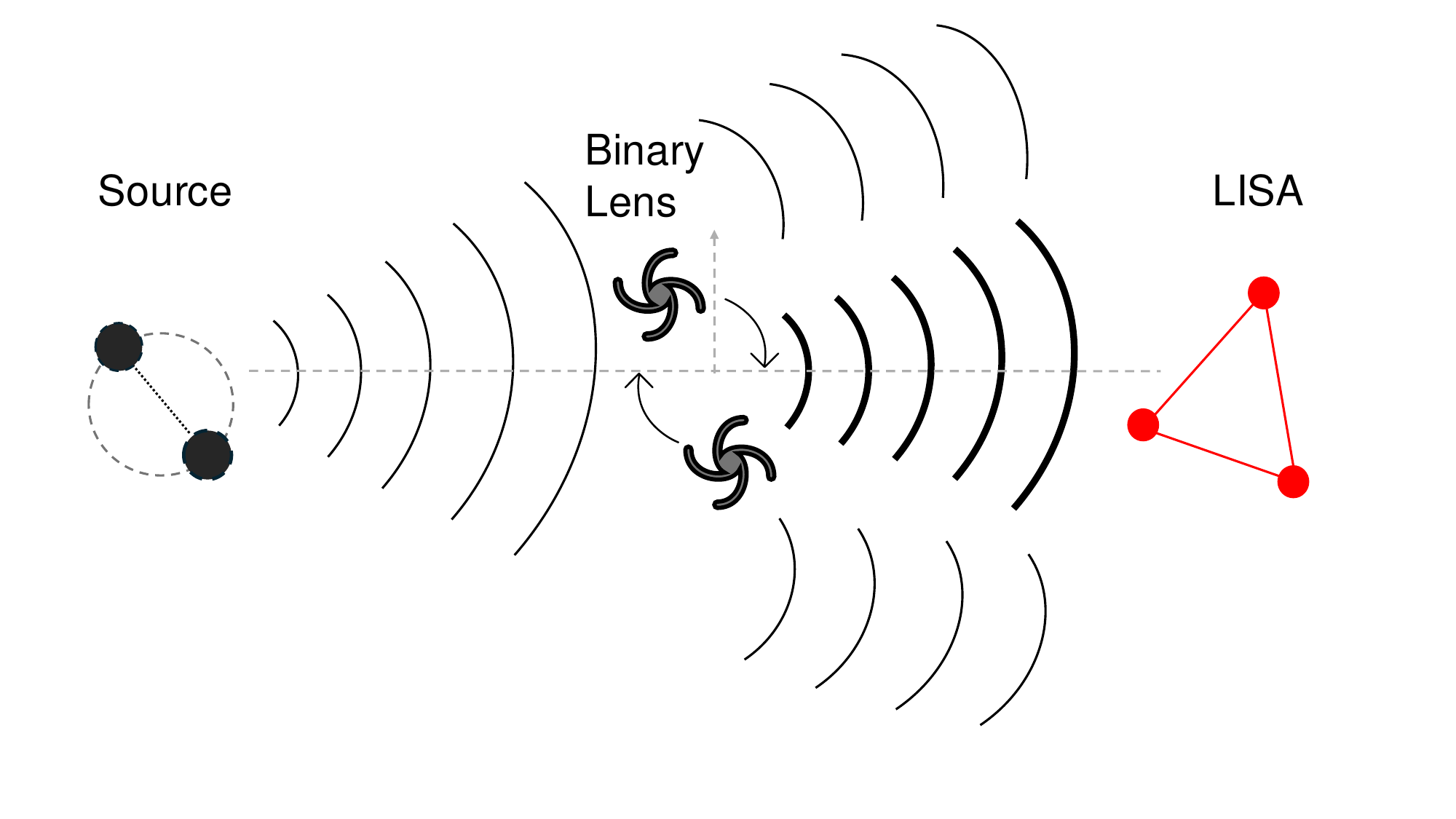}
    \caption{Schematic of the gravitational wave source, binary black hole lens system, and LISA detector, emphasizing periodic amplitude modulation (AM) effects induced by the dynamic lens system \cite{qiu2022}.}
    \label{fig:AM}
\end{figure}

This modulation mechanism enables gravitational wave detectors, such as LISA, to capture these gravitational wave characteristics within their sensitive frequency ranges (illustrated in Fig.~\ref{fig:AM}), offering a novel approach for detecting binary black hole lensing systems \cite{haris2018}, \cite{phinney2002}. 

\subsection{Frequency Modulation Based on Dark Matter}

While amplitude modulation leverages astrophysical phenomena, frequency modulation explores the influence of dark matter, uncovering additional mechanisms for wave manipulation.

Frequency modulation (FM) of gravitational waves involves the presence of ultralight scalar dark matter (ULDM) and its impact on the local gravitational potential along the gravitational wave propagation path. \cite{wang2023} explores how the oscillating pressure of ULDM causes periodic oscillations in the local gravitational potential. Gravitational waves passing through these regions experience changes similar to the time-dependent frequency drift observed in pulsar signals. 

This effect is quantified through the frequency shift formula: the observed frequency shift includes not only the time-independent part \(\Psi_{c}(x_s)\) but also the time-varying part \(\Psi_o(x_s) \cos(\omega t)\). The time-independent part causes an overall frequency redshift, while the time-varying part induces periodic frequency modulation, leading to the gravitational wave signal's frequency oscillating up and down over the observation period. To enhance this effect, researchers assume that white dwarf binary systems are located in dark matter clumps or subhalos with extremely high dark matter density, allowing the minute frequency drifts to be captured by high-sensitivity gravitational wave detectors like LISA. Analysis using the Fisher information matrix indicates that LISA can detect frequency modulations when the ULDM mass is in the range of \( 1.67 \times 10^{-23} \sim 4.31 \times 10^{-23} \, \text{eV}/c^{2} \), LISA can detect these frequency modulations. 

The study indicates that if white dwarf binary systems are located within dark matter clumps, their gravitational wave signals exhibit significant frequency modulation during the LISA mission period. This finding not only provides a potential modulation mechanism for future gravitational wave communication but also offers a new window for the indirect detection of ULDM. 

\subsection{Manipulating Gravitational Waves via Superconductors}

Transitioning from astrophysical to laboratory-based modulation methods, manipulating gravitational waves using superconductors presents an intriguing possibility. Based on the theoretical predictions in \cite{li1992}, \cite{torr1993}, the phase velocity of gravitational waves in superconductors decreases by about 300 times, and the wavelength shortens to 1/300 of that in free space. This phenomenon enables the manipulation of gravitational waves within superconductors, including focusing, reflection, and refraction, laying the foundation for implementing phase and frequency modulation of gravitational waves. 

\begin{figure}[!t]
    \centering
    \includegraphics[width=\linewidth]{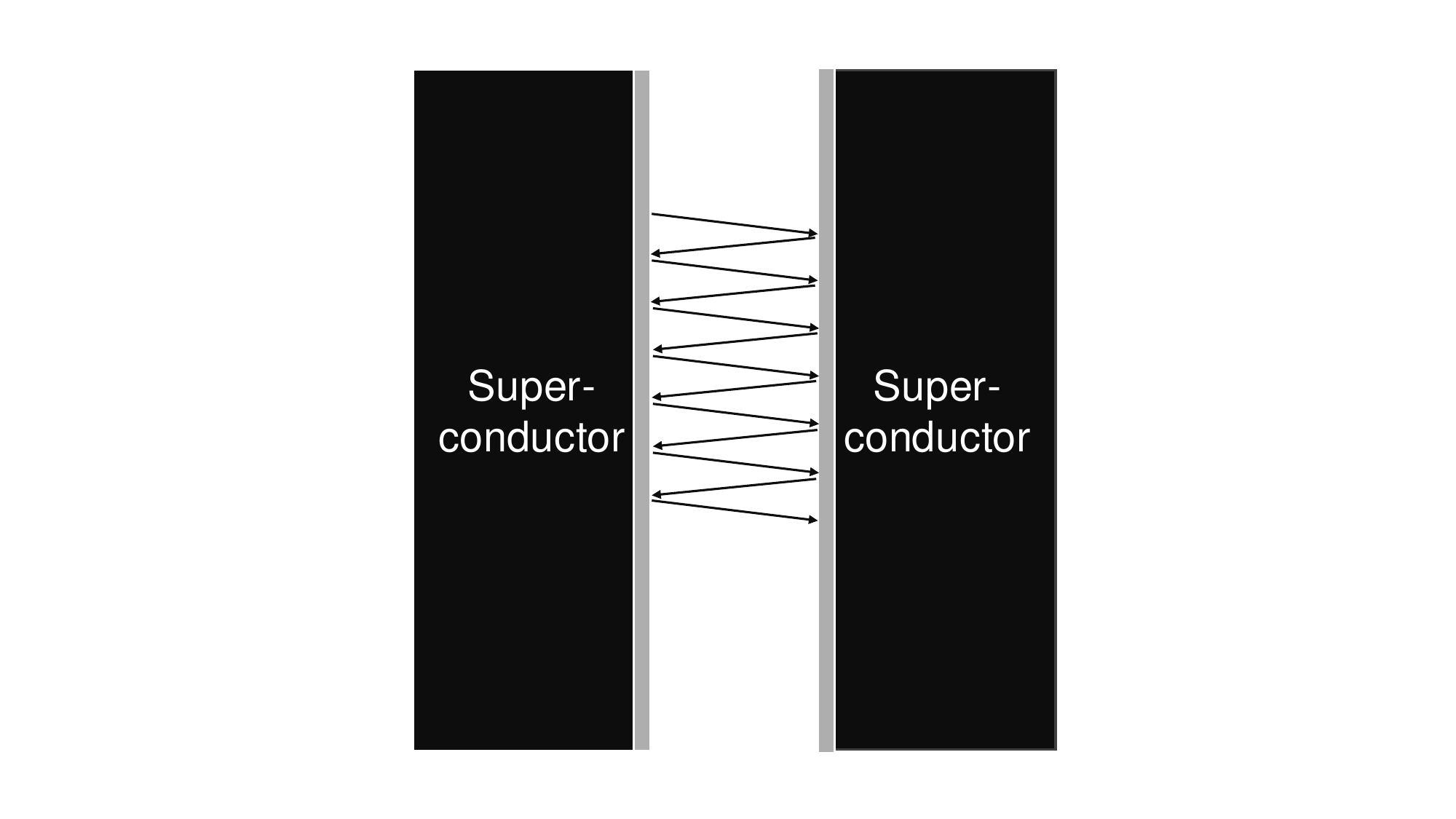}
    \caption{Fresnel reflection of HFGW flux at the superconductor-air interface, with a high reflection coefficient and low transmission coefficient. These properties enable Fabry-Pérot resonator-like structures for gravitational wave frequency modulation.}
    \label{fig:mod_c}
\end{figure}

\cite{woods2005} further discuss the Fresnel reflection of gravitational waves at the superconductor-air interface. Due to the significant difference in refractive indices between superconductors and air, the reflection coefficient is close to 1, and the transmission coefficient is only 1.3\%. Using these high-reflection and low-transmission characteristics, structures similar to optical Fabry-Pérot resonators can be designed as illustrated in Fig.~\ref{fig:mod_c}. By adjusting the distance between reflective surfaces, resonance effects at specific frequencies can be achieved, realizing frequency modulation of gravitational waves. 

\subsection{Gravitational Wave Modulation Based on Nonmetricity}

Moving from an experimental modulation approach based on material properties to a more theoretical framework, we explore the potential of non-metric modulation, which expands the scope of gravitational wave communication to include new geometric properties of spacetime. \cite{babourova2018} utilize Lie derivatives to derive the components of nonmetricity waves. They discovered that these gravitational waves are determined by arbitrary functions of delayed time, meaning that nonmetricity waves can be modulated by selecting different arbitrary functions to carry and transmit information, similar to the modulation process in electromagnetic waves. 

This type of gravitational wave differs from those under the traditional general relativity framework, as it considers nonmetricity—a geometric property where the metric tensor does not remain invariant during parallel transport—in extended gravitational theories \cite{quiros2022}. The study of nonmetricity gravitational waves expands the theoretical scope of gravitational waves, providing a new medium for information transmission. 


\subsection{Open Research Challenges}

Despite the diverse possibilities offered by gravitational wave modulation techniques, several significant challenges remain.

\subsubsection{Practical Detection and Parameter Extraction} Astrophysical modulation methods like amplitude modulation are theoretically feasible but practically challenging. Low-frequency signals are hard to detect with current instruments like LISA, and extracting parameters such as mass ratios from modulated signals remains difficult. Progress requires innovative detection and signal processing methods, possibly adapting radio communication techniques.

\subsubsection{Challenges in Communication Theory}
Despite the fundamental applicability of conventional communication theory to gravitational wave signals, the ultra-low SNR, non-Gaussian and non-stationary noise, and scarcity of controllable sources create significant obstacles. For instance, proven frameworks like matched filtering or error-correcting codes struggle to achieve robust performance in these extreme conditions, while potential extensions of multi-antenna (or multi-detector) methods remain speculative without large-scale gravitational wave sensor arrays. Adapting standard modulation and coding strategies from electromagnetic communication to this unorthodox medium demands further research that accounts for unique constraints such as quantum noise, limited bandwidth, and the inherent difficulty of generating and shaping gravitational waves. Cross-disciplinary collaboration among physicists, communication engineers, and astronomers is essential to bridge these theoretical gaps and guide any future proof-of-concept demonstrations.

\subsubsection{Uncertainties in Dark Matter Properties and Distribution} Frequency modulation involving ultralight scalar dark matter (ULDM) depends on uncertain assumptions about dark matter's properties and distribution. Effective modulation requires white dwarf binaries within dense dark matter clumps, but their actual distribution and ULDM characteristics are not well understood. Resolving these uncertainties is crucial for advancing gravitational wave communication and dark matter research.

\subsubsection{Manufacturing Precision and Energy Loss} Laboratory methods using superconductors to manipulate gravitational waves face practical issues like extreme manufacturing precision and significant energy loss. Short wavelengths demand machining tolerances within ±20 micrometers for 4.9 GHz waves, and high reflectivity at interfaces causes energy loss, reducing efficiency. Future research should focus on new materials or structures to mitigate energy loss, improve precision manufacturing, and design more efficient devices.

\subsubsection{Experimental Verification of Nonmetricity Gravitational Waves} Nonmetricity-based gravitational wave modulation offers a novel theoretical framework but lacks experimental validation. The generation, propagation, and detection mechanisms are not fully understood. Further research is needed to verify their existence and explore practical communication applications.

\subsubsection{Impact of Cosmic Environments on Gravitational Wave Propagation} Interstellar propagation of modulated gravitational waves faces challenges from cosmic environments. Factors like magnetic fields and cosmic perturbations can cause attenuation and noise, reducing communication reliability. These factors can lower the signal-to-noise ratio and complicate detection. Addressing these challenges requires a deeper understanding of gravitational wave interactions with cosmic environments. The next section will discuss these channel components, analyzing their dual impact on gravitational wave transmission and implications for the future of this technology.

\section{Gravitational Wave Communication Channel}

\begin{figure*}[!t]
    \centering
    \includegraphics[width=\linewidth]{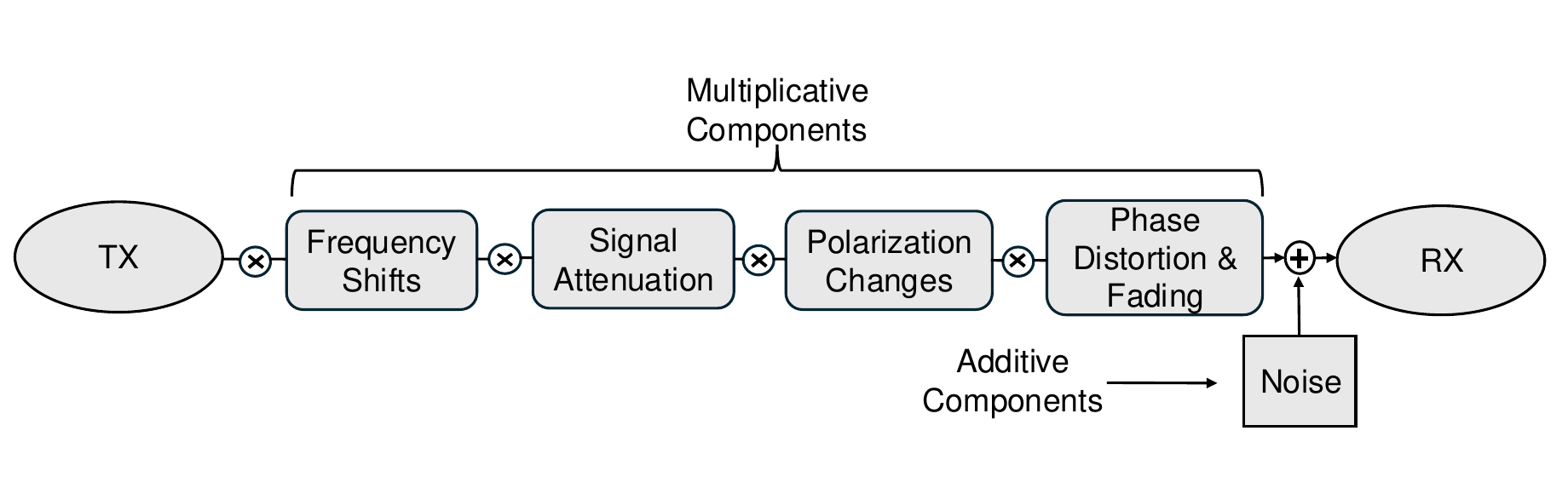}
    \caption{A conceptual illustration of the gravitational wave communication channel architecture, arranged according to the physical origins and scales of different effects. The signal first experiences large-scale influences such as gravitational and cosmological frequency shifts, followed by broad-scale amplitude attenuation due to cosmic expansion and weak scattering. Next, more region-specific factors induce polarization changes, and finally, localized distortions arise in the form of phase variations and fading effects caused by gravitational lensing and other fine-scale phenomena. Additive noise is introduced near the receiver end. This hierarchical arrangement reflects the underlying physics and facilitates clearer analytical modeling for communication system design.}
    \label{fig:channel}
\end{figure*}

In electromagnetic (EM) communication, a channel is typically defined by the propagation environment, such as wired or wireless media, each with specific compositions and impacts on signal transmission \cite{sambo2014}. Similarly, in gravitational wave communication, the channel refers to the propagation path of gravitational waves, shaped by cosmic structures and fields that influence signal properties, including attenuation, frequency shifts, phase distortion, fading effects, polarization changes, and noise components. The sequence in which these effects are considered can be guided by their physical origins and scales. For instance, frequency shifts like gravitational or cosmological redshifts are driven by large-scale spacetime structures—such as cosmic expansion or deep gravitational potential wells—and naturally arise before more localized phenomena. Next, overall amplitude reduction due to broad-scale influences (e.g., cosmic expansion, weak scattering by interstellar media) addresses the global impact on signal power. Following these macro-level effects, polarization changes, often associated with more region-specific fields (such as local magnetic or electric fields, or asymmetric spacetime geometries), come into play, altering the wave’s polarization state after the initial frequency and amplitude adjustments have been accounted for. Finally, phase distortions and fading—stemming from relatively localized and subtle factors like small-scale curvature variations, gravitational lensing, and multipath interference near massive objects—occur later in the propagation chain. Thus, the channel components can be conceptually arranged from large-scale, broadly acting factors to increasingly localized and fine-grained effects, with additive noise introduced near the receiving end. This hierarchical ordering reflects the underlying physics and allows for clearer analytical modeling and simulation. Figure~\ref{fig:channel} provides a visual representation of this channel architecture and the relative positions of each component. This section explores these gravitational wave communication channel components in detail, highlighting their roles in shaping gravitational wave propagation and their implications for developing effective communication systems.

\subsection{Frequency Shifts}

Frequency shifts involve changes in the frequency of gravitational waves as they traverse varying gravitational potentials and cosmic structures. The primary causes of frequency shifts are:

\subsubsection{Time Dilation and Gravitational Redshift} In regions of intense gravitational fields, such as near massive objects like black holes or neutron stars, gravitational waves experience a gravitational redshift. This effect arises due to universe's expansion and time dilation, where time slows down relative to an external observer. The curvature of spacetime in these regions stretches the waves, lowering their frequency as observed from a distant location. This phenomenon plays a significant role in shaping the characteristics of gravitational waves emitted in strong-field environments \cite{palessandro2020}, \cite{fier2021}. 

\subsubsection{Cosmological Redshift} On cosmological scales, the expansion of the universe stretches the wavelength of gravitational waves over vast distances, causing a shift to lower frequencies. This phenomenon, known as the cosmological redshift, is particularly significant for emissions originating in the early universe, such as those produced by primordial black hole evaporation or other high-energy astrophysical events. In the matter-dominated universe after recombination, the cosmological redshift \( z \) reduces the observed energy \( E_{m,0} \) of the emitted particles, and is expressed as:

\begin{equation}
    E_{m,0} = \frac{E_m}{1 + z} \label{eq:4},
\end{equation}
where \( E_{m,0} \) is the observed energy of the emitted particles at the present time, \( E_m \) is the energy of the emitted particles at the time of emission, and \( z \) is the cosmological redshift, which quantifies the stretching of wavelengths due to the universe's expansion. In this era, the redshift \( z \) is related to cosmic time \( t \) by the relation:

\begin{equation}
    1 + z = \left( \frac{5 \cdot 10^{17}}{t} \right)^{2/3} \label{eq:5}.
\end{equation}

Substituting~\eqref{eq:5} into~\eqref{eq:4}, the observed energy of particles emitted from primordial black hole evaporation becomes:

\begin{equation}
    E_{m,0} = 105 \cdot \left( \frac{M}{4 \cdot 10^{14}} \right)^{-1} \, \text{MeV},
\end{equation}
where \( M \) is the initial mass of the evaporating primordial black hole. The energy \( E_{m,0} \) observed today is therefore inversely proportional to the initial black hole mass and decreases due to the cosmological redshift.

Before recombination, in the radiation-dominated universe, the evolution of the cosmological redshift follows a different relation. In this epoch, the redshift \( z \) is related to the cosmic time \( t \), the recombination redshift \( z_{\text{rec}} \), and the recombination time \( t_{\text{rec}} \) as:

\begin{equation}
    1 + z = \left( \frac{t_{\text{rec}}}{t} \right)^{1/2} \cdot (1 + z_{\text{rec}}),
\end{equation}
where \( t_{\text{rec}} \) is the cosmic time at recombination, and \( z_{\text{rec}} \) is the redshift at the recombination epoch. The energy of the emitted particles at recombination \( E_{m,\text{rec}} \) can be related to the energy at the time of emission \( E_m \) through the expression:

\begin{equation}
    E_{m,\text{rec}} = E_m \cdot \frac{1 + z_{\text{rec}}}{1 + z}.
\end{equation}

These relations demonstrate that the expansion of the universe has a profound effect on the frequency and energy of gravitational waves emitted in the early universe. High-frequency emissions from primordial black hole evaporation or similar processes experience significant redshift over billions of years, transitioning to lower frequencies detectable by modern gravitational wave instruments \cite{bisnovatyi2004}.


\subsubsection{Interrelationships with Other Components} Frequency shifts often occur alongside attenuation in intense gravitational fields. Both are consequences of the same underlying phenomena—time dilation and energy absorption. These frequency shifts complicate signal demodulation and extraction, requiring precise models and compensation algorithms to correct for these changes.

\subsection{Signal Attenuation}

Signal attenuation refers to the reduction in amplitude or strength of the gravitational wave signal as it propagates through various cosmic environments. This phenomenon is influenced by several factors:

\subsubsection{Intense Gravitational Fields} In the vicinity of dense astrophysical objects, such as black holes and high-density celestial bodies, gravitational waves may be affected to some extent. For instance, near a black hole’s event horizon, the strong gravitational field could absorb part of the wave’s energy, leading to a slight reduction in signal amplitude. However, \cite{palessandro2020} emphasizes that under the framework of Einstein gravity and minimally coupled matter, the absorption of gravitational waves is highly inefficient, making this energy absorption negligible in most scenarios.

This conclusion is quantitatively supported by the gravitational absorption cross-section, expressed as:
\begin{equation}
    \sigma_{\text{abs}} = \frac{3^4 \pi^2}{5 \times 2^9} l_p^2 \approx 0.3 l_p^2,
\end{equation}
where $l_p$ is the Planck length. This formula indicates that the gravitational absorption cross-section is universal and independent of the mass or coupling of the absorbing bound state. With a value comparable to the Planck area, it is extremely small. As a result, even in the proximity of dense celestial objects or black holes, the absorption of gravitational waves remains negligible. Furthermore, dense celestial bodies also cause amplitude reduction, especially at low frequencies, due to gravitational absorption.

\subsubsection{Interstellar Medium and Dust} Gravitational waves interacting with interstellar dust and gas may undergo slight scattering and absorption, resulting in limited attenuation over extremely long distances. \cite{palessandro2020} highlights that the cumulative effects of such interactions could potentially become significant on a cosmic scale. However, similar to dense astrophysical regions, the absorption remains weak due to the extremely small cross-section, generally insufficient to cause notable changes in the gravitational wave signals \cite{bartelmann2001}. 

\subsubsection{Magnetic Fields and Plasma Environments} In magnetized plasma environments, such as those found near magnetars or within galactic cores, gravitational waves may partially convert into electromagnetic waves. This conversion leads to energy loss and signal attenuation \cite{bisnovatyi2004}. Moreover, the large-scale but weak magnetic fields of galaxy clusters can cumulatively attenuate gravitational waves due to their vast spatial extent \cite{bamba2018}.

\subsubsection{Complex Geometric Structures} Variations in spacetime curvature, influenced by the presence of magnetic or electric fields, lead to significant changes in the propagation characteristics of gravitational waves. These effects are explicitly calculated in the context of cylindrically symmetric, axially symmetric, and spherically symmetric spacetime structure \cite{barrabes2010}. 

For example, in the spherically symmetric case, the interaction with a weak magnetic field introduces additional modifications to the wavefront geometry, as described by the Weyl tensor component:

\begin{equation}
C_{2323} - iC_{2324} = \left(\frac{1}{r} G^{-1} F(\zeta) + G(\zeta)\right) \delta(u),
\end{equation}
where $C_{2323}$ and $C_{2324}$ are components of the Weyl tensor, describing the amplitude and geometric distortions of the gravitational wave caused by spacetime curvature and electromagnetic interactions and $F(\zeta)$ and $G(\zeta)$ represent analytic functions encoding the influence of the magnetic field. These modifications highlight how spacetime curvature and background electromagnetic fields alter the wave's amplitude and geometry.

Such effects highlight the role of spacetime geometry in modifying gravitational wave propagation and introducing signal attenuation in channel.

\subsubsection{Atypical Cosmic Structures} In certain cosmological models, such as K-matter-dominated universes with linear expansion rates, gravitational waves experience continuous attenuation without any restoration mechanisms \cite{mondal2021}. The ongoing expansion of space itself stretches the gravitational waves, reducing their amplitude over time.

\subsubsection{Interrelationships with Other Components} Attenuation is often accompanied by frequency shifts due to gravitational redshift and time dilation in intense gravitational fields. Additionally, energy loss from attenuation can affect the phase of the wave, leading to phase distortions. These combined effects present significant challenges for signal detection and interpretation.

\subsection{Polarization Changes}

Polarization changes involve alterations in the orientation or state of polarization of gravitational waves during propagation. These changes affect the ability to accurately detect and interpret the signal. The main causes are:

\subsubsection{Strong Gravitational Fields} Near massive objects like black holes, the extreme curvature of spacetime can significantly influence the polarization characteristics of gravitational waves. These effects include alterations in the polarization state due to the interaction with the curved spacetime geometry and potential spin effects of the lensing body. The polarization content, often assumed to be parallel transported along the propagation path in the geometric optics limit, may exhibit additional modifications in the wave optics regime. Such changes could accumulate over long propagation distances and depend on the specific spacetime geometry and the nature of the lensing object. \cite{braga2024}.


\subsubsection{Anisotropic Universes} In cosmological models like the anisotropic Kasner universe, directional expansion rates cause different polarization effects depending on the wave's propagation direction. This anisotropy leads to variations in attenuation and frequency shifts, complicating the polarization characteristics \cite{mondal2021}.

\subsubsection{Impact on Communication} Polarization changes present challenges for signal demodulation, as receivers must be capable of detecting and processing multiple polarization modes. Variations in polarization contribute to amplitude fluctuations, affecting signal stability and requiring advanced polarization-handling techniques in receiver design.

\subsection{Phase Distortion}

Phase distortion refers to changes in the phase of gravitational waves, affecting the waveform and timing of the signal. This distortion arises from several factors:

\subsubsection{Gravitational Lensing} Massive objects like black holes and galaxies can bend spacetime, altering the propagation path of gravitational waves. This bending causes phase delays and distortions as the waves take longer paths or traverse regions of different gravitational potentials \cite{braga2024}. This effect can be quantitatively described by the amplification factor equation:

\begin{equation}
\frac{1}{r^2} \nabla^2_\theta F + \frac{\partial^2 F}{\partial r^2} + 2i\omega \frac{\partial F}{\partial r} = 4\alpha \omega^2 U F
\end{equation}
where \( F \)  represents the amplification factor that encodes the wave’s phase and amplitude changes, and \( U \)  is the gravitational potential of the lensing object. The term \( 4\alpha \omega^2 U F \) captures the influence of the gravitational potential on the wave propagation, leading to phase delays and distortions as the waves take longer paths or traverse regions of different gravitational potentials. Gravitational lensing can lead to multipath propagation, where waves traveling along different paths experience varying phase shifts, resulting in observable phase distortions at the receiver.

\subsubsection{Magnetic Fields} Strong magnetic fields, especially in regions near magnetars or galactic centers, can induce phase shifts in gravitational waves \cite{bamba2018}. Interactions between gravitational waves and magnetic fields modify the wave's phase due to coupling effects.


\subsubsection{Interrelationships with Other Components} Phase distortion contributes to multipath fading by causing the superposition of signals with varying phases. It can also be accompanied by signal attenuation, further affecting overall signal integrity and complicating signal processing efforts.

\subsection{Fading Effects}

Fading refers to fluctuations in signal amplitude and phase over time or space, impacting signal consistency and reliability. Different types of fading effects in gravitational wave communication include:

\subsubsection{Multipath Fading} Caused by gravitational lensing, multipath fading occurs when gravitational waves propagate along multiple paths with varying delays and phases due to the curvature of spacetime around massive objects \cite{braga2024}. This effect arises as the gravitational potential of the lensing object bends the wavefronts, forcing the waves to traverse different geometric and gravitationally delayed paths. The resulting amplification factor \( F \), under the thin lens approximation, is given by:  
\begin{equation}
F(\vec{r}_O) = \int d^2\theta_L \exp \left\{ i\omega \left[ \frac{r_L r_O}{2 r_{LO}} |\theta_L - \theta_O|^2 - \hat{\psi}(\theta_L) \right] \right\},  
\label{eq:amplification_factor}
\end{equation}
where \( \hat{\psi}(\theta_L) \) represents the gravitational potential of the lens, and the term \( \frac{r_L r_S}{2 r_{LS}} |\theta_L - \theta_O|^2 \) accounts for the geometric time delays caused by varying path lengths. The convergence of these paths at the receiver results in constructive and destructive interference, leading to periodic fluctuations in the signal's amplitude and phase. This interference pattern quantifies the multipath fading phenomenon, where the signal stability is affected by the relative phase differences and delays introduced by the curvature of spacetime.

\subsubsection{Polarization-Related Fading} Changes in the polarization states of gravitational waves, induced by strong gravitational or magnetic fields, can lead to mismatches between the transmitted and received polarization modes. This mismatch causes amplitude variations and contributes to fading effects similar to polarization fading in electromagnetic communications \cite{bamba2018}.

\subsubsection{Magnetar-Induced Fading} Intense and non-uniform magnetic fields around magnetars create multipath interference and signal fading. The spatial non-uniformity of these strong magnetic fields causes periodic amplitude changes and may result in intermittent signal loss, posing challenges for signal detection and stability \cite{duncan2003}, \cite{bamba2018}.

\subsubsection{Interrelationships with Noise Components} Fading effects can introduce coherent noise due to predictable interference patterns. The presence of fading increases the complexity of signal extraction and necessitates sophisticated algorithms to mitigate its impact on signal quality.

\subsection{Noise Components}

Noise components are unwanted disturbances that interfere with the gravitational wave signal, reducing the signal-to-noise ratio (SNR) and complicating signal extraction. The types of noise in a gravitational communication channel include:

\subsubsection{Thermal Gravitational Noise} In high-temperature plasma environments, such as stellar cores or supernova envelopes, the thermal motion of particles introduces random noise components. This noise overlaps with the gravitational wave signal, reducing SNR and increasing the difficulty of signal analysis \cite{bisnovatyi2004}.

\subsubsection{Coherent Noise} Predictable interference patterns arising from multipath propagation and gravitational lensing contribute to coherent noise. When gravitational waves traverse curved spacetime around massive objects, the bending of trajectories creates multiple propagation paths. These paths accumulate different phase delays due to varying lengths and gravitational potentials, leading to interference effects \cite{braga2024}. The diffraction integral describing this phenomenon is given as:
\begin{equation}
F(\vec{r}_O) = \int D\theta(r) \exp\left( i\omega \int_0^{r_O} dr L[\theta(r), \dot{\theta}(r), r] \right),
\end{equation}
where $F(r_O)$ is the amplification factor, and $L$ represents the accumulated phase along each path. 

In the high-frequency limit ($\omega \to \infty$), the integral is dominated by stationary points of the time delay function, corresponding to the classical ray trajectories. However, at lower frequencies, multiple paths contribute to the final amplitude, resulting in interference patterns that can be observed as coherent noise in gravitational wave detection.

\subsubsection{Background Gravitational Noise} Low-frequency disturbances from astrophysical sources and the interstellar medium accumulate over vast distances. This background noise increases the complexity of signal extraction and can obscure weaker gravitational wave signals \cite{palessandro2020}.

\subsubsection{Quantum Interference Noise} In scenarios involving extreme bound states or interactions with quantum fields, quantum interference effects emerge as the overlap of wavefunctions during transitions between quantum states. These effects, particularly in high-density or near-collapse conditions, influence the absorption and response characteristics of gravitational waves. While these interactions do not inherently introduce noise, they highlight the universal and quantized nature of the gravitational wave absorption process. \cite{palessandro2020}.

\subsubsection{Interrelationships with Other Components} Noise components can exacerbate fading effects and may mask or mimic attenuation and phase distortions. Advanced signal processing techniques are required to separate noise from the desired signal and enhance overall detection capabilities.

\subsection{Propagation in Different Environments}

Gravitational waves propagate through various cosmic environments, each presenting unique challenges and effects on the signal. Understanding the propagation characteristics in different environments is essential for designing reliable gravitational communication systems.

\subsubsection{Propagation in High-Density and Extreme Gravitational Fields}

\begin{itemize}
    \item \textbf{Near Black Holes:} In the vicinity of black holes, gravitational waves are profoundly affected by intense gravitational forces. Signal attenuation occurs due to energy absorption by the strong gravitational field near the event horizon \cite{palessandro2020}. Time dilation and gravitational redshift decrease the signal frequency and alter the phase. The super-radiance effect in rotating black holes can amplify signals at specific frequencies, altering the spectral characteristics of the gravitational wave \cite{fier2021}. Gravitational lensing causes multipath propagation, leading to phase distortions and multipath fading \cite{braga2024}.

    \item \textbf{Dense Celestial Bodies and Plasma Environments:} When gravitational waves pass through high-density regions like neutron stars or dense plasma environments, their amplitude decreases, especially at low frequencies. Thermal gravitational noise from high-temperature plasma introduces random noise components, reducing SNR and complicating signal extraction \cite{bisnovatyi2004}.
\end{itemize}

\subsubsection{Propagation in Magnetic Fields and Plasma Environments}
\begin{itemize}
    \item \textbf{Magnetars and Galactic Cores:} Strong magnetic fields in these regions cause gravitational Faraday rotation, affecting the polarization of gravitational waves \cite{palessandro2020}, \cite{fier2021}. Non-minimal coupling between gravitational waves and magnetic fields leads to slight changes in amplitude and polarization. Phase shifts are induced by interactions with magnetic fields \cite{bamba2018}, and signal attenuation occurs due to energy loss in non-minimal coupling scenarios \cite{bisnovatyi2004}.

    \item \textbf{Interstellar Magnetized Plasma:} Widespread in star-forming regions and galactic environments, magnetized plasma affects gravitational wave propagation by partially converting gravitational waves into electromagnetic waves, leading to signal attenuation and polarization changes. The cumulative effects of large-scale but weak magnetic fields can significantly attenuate the signal over long distances \cite{bamba2018}.

    \item \textbf{Galaxy Clusters and Weak Magnetic Fields:} The magnetic fields within galaxy clusters, though relatively weak, have cumulative effects on gravitational wave signals due to their vast spatial scale. As gravitational waves propagate through these large-scale, low-intensity magnetic fields, their amplitude gradually diminishes \cite{bamba2018}.

    \item \textbf{High-Magnetic-Field Regions at Galaxy Centers:} In regions near active black holes, magnetic field strength significantly intensifies. The complex spatial distribution of magnetic fields here has a substantial impact on gravitational wave propagation, inducing phase shifts and causing variations in polarization components \cite{bamba2018}.
    
\end{itemize}

\subsubsection{Propagation through Interstellar Medium, Dust, and Radiation}
\begin{itemize}
    \item \textbf{Interstellar Medium and Dust:} Gravitational waves propagating through the interstellar medium and cosmic dust experience slight attenuation and scattering. Over ultra-long distances, these cumulative effects become significant, leading to minor decreases in signal amplitude and slight deflections in the propagation path. Gravitational lensing and dust distribution cause minor signal delays and phase disruptions, reducing signal clarity \cite{palessandro2020}, \cite{bartelmann2001}.

    \item \textbf{Cosmic Microwave Background (CMB) Radiation:} The CMB radiation has a slight attenuating effect on gravitational waves. Weak interactions with interstellar particles, dust, and small particles in regions rich with these components lead to energy loss and further signal attenuation \cite{bisnovatyi2004}.

    \item \textbf{Background Gravitational Noise:} The accumulation of low-frequency disturbances from astrophysical sources and the sparse gas and dust in the intergalactic medium introduces background gravitational noise. This noise increases the complexity of signal extraction and may mask weaker signal components \cite{palessandro2020}.
\end{itemize}

\subsubsection{Propagation in Atypical Cosmic Structures}
\begin{figure*}[!t] 

    \centering 

    \includegraphics[width=\linewidth]{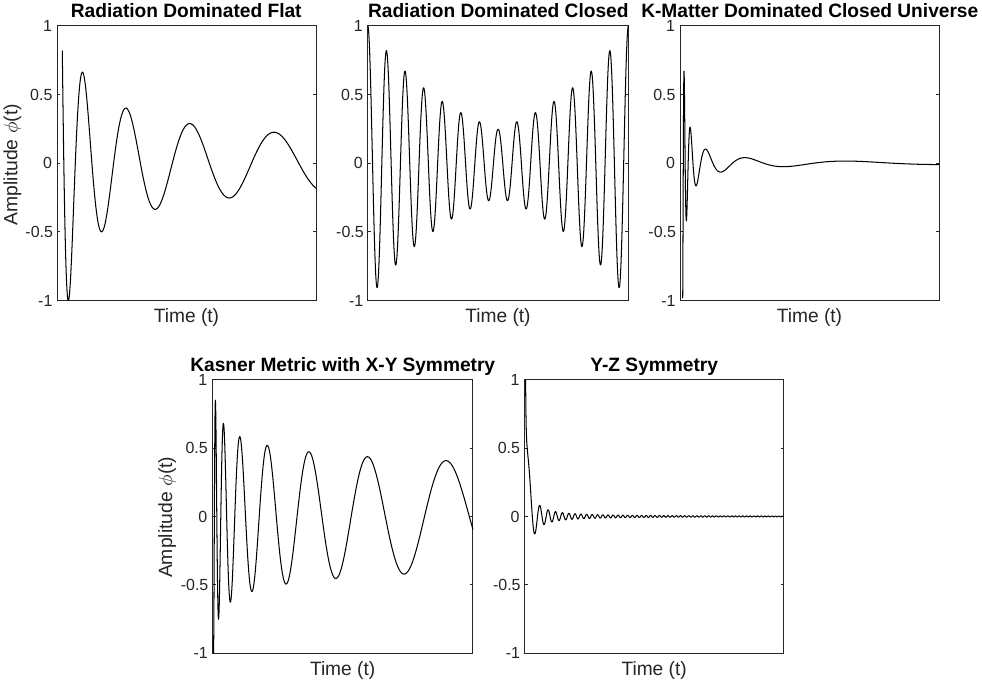} 

    \caption{Comparative amplitude variations of gravitational waves propagating through four atypical cosmic structures. The radiation-dominated flat universe demonstrates gradual amplitude decay due to slow cosmic expansion, while the radiation-dominated closed universe exhibits significant signal recovery during the contraction phase. In the K-matter-dominated closed universe, continuous signal attenuation occurs without restoration, reflecting a linear expansion rate. The anisotropic Kasner universe introduces directional effects, with varying attenuation and frequency shifts depending on propagation direction. These visualizations highlight the distinct propagation characteristics of gravitational waves under different cosmological backgrounds.} 

    \label{fig:atypical} 

\end{figure*} 

\begin{itemize}
    \item \textbf{Radiation-Dominated Flat FRW Universe:} In a radiation-dominated flat Friedman-Robertson-Walker (FRW) universe, gravitational waves experience gentle amplitude decay and frequency drops due to slow cosmic expansion. The expanding background causes gradual weakening, allowing signals to propagate more stably over greater distances \cite{ahmed2019}, \cite{cai2003}.

    \item \textbf{Radiation-Dominated Closed FRW Universe:} This universe gradually contracts after reaching a certain expansion limit, leading to a ``Big Crunch." During the contraction phase, gravitational wave amplitudes and frequencies gradually recover, reaching an enhanced state \cite{grishchuk1981}, \cite{caldwell1993}. This recovery effect creates a signal peak, potentially causing dramatic volatility in signal detection.

    \item \textbf{K-Matter-Dominated Closed FRW Universe:} In this model, the universe has a constant expansion rate, resulting in continuous signal attenuation and frequency reduction without restoration \cite{mondal2021}. The lack of a restoration mechanism limits long-distance propagation and may lead to complete signal loss over greater distances.

    \item \textbf{Anisotropic Kasner Universe:} The Kasner universe is a spatially flat but anisotropic universe with different expansion rates in three directions. Gravitational wave propagation exhibits directional effects, with attenuation and frequency shifts varying based on the propagation direction. This anisotropy leads to complex polarization characteristics and requires consideration in communication system design \cite{rodrigues2008}.
    
\end{itemize}

In summary, these atypical cosmic structures demonstrate a range of gravitational wave propagation characteristics, including gradual attenuation and restoration effects in radiation-dominated universes, continuous signal loss in K-matter-dominated universes, and directional effects in the Kasner universe. To illustrate these propagation patterns, Fig.~\ref{fig:atypical} visualizes how gravitational waves propagate differently in these cosmic structures, highlighting their unique amplitude variations across different atypical cosmic expansion. This figure highlights the contrasting signal behaviors, from the gradual amplitude decay in radiation-dominated flat universes to the dramatic signal recovery in radiation-dominated closed universes, the continuous weakening in K-matter-dominated universes, and the anisotropic directional effects in Kasner metrics. Such visual comparisons offer valuable insights into the complex wave dynamics in these unique cosmological backgrounds, thereby providing theoretical support for optimizing gravitational communication system designs. 

\subsubsection{Interrelationships with Other Components}

Understanding the propagation of gravitational waves in different environments highlights the interrelationships between channel components. For instance, significant attenuation and frequency shifts in intense gravitational fields are intrinsically linked to time dilation and gravitational redshift. Phase distortion and multipath fading arise from gravitational lensing, causing multiple paths with varying phases and leading to interference patterns.

\subsection{Open Research Challenges}
Despite the comprehensive analysis of various cosmic structures and their effects on gravitational communication channels, several critical areas remain underexplored. Addressing these open issues is crucial for advancing the field and achieving reliable and efficient gravitational communication. Additionally, these unresolved challenges have significant implications for system design, signal processing, and overall communication performance.

\subsubsection{Interaction Between Gravitational Waves and Quantum Fields}
At extremely small scales, the interaction between gravitational waves and quantum fields could have non-negligible effects on signal propagation. Quantum fluctuations and vacuum polarization may influence gravitational waves, affecting channel reliability and accuracy. Understanding these interactions requires integrating principles from both general relativity and quantum mechanics \cite{palessandro2020}. For communication systems, the potential quantum-induced variations necessitate robust signal processing techniques that can adapt to stochastic effects.

\subsubsection{Lack of a Unified Theory of Quantum Gravity}
General relativity and quantum mechanics remain fundamentally incompatible, and the quantization of gravity is one of the most challenging problems in theoretical physics. Without a unified quantum gravity framework, the behavior of gravitational waves at quantum scales remains uncertain. This theoretical gap limits the ability to predict or simulate quantum-scale impacts on gravitational wave channels, complicating both experimental studies and practical applications.

\subsubsection{Propagation Characteristics at Quantum Scales}The propagation of gravitational waves through quantum dominated environments remains poorly understood. Potential coupling effects with other quantum fields, such as scalar or vector fields, and phenomena predicted by quantum gravity theories, could introduce novel impacts on gravitational communication channels \cite{palessandro2020}. These effects might influence signal attenuation, phase distortion, and noise characteristics in ways not accounted for by classical models.

\subsubsection{Information-Theoretic Challenges in Gravitational Communication}

Although information theory principles are, in principle, applicable to any physical transmission medium, gravitational communication channels pose unique theoretical and practical hurdles. From a classical standpoint, Shannon’s framework allows us to characterize channel capacity by analyzing factors such as signal power, bandwidth, and noise \cite{shannon1948}. However, gravitational waves are typically subject to extreme conditions: severely low signal-to-noise ratios (SNR), non-Gaussian and non-stationary noise sources, and highly limited control over the transmission process. These factors may drive the channel capacity to extremely small values or even render it effectively zero. Nonetheless, performing a rigorous capacity analysis remains an open question, as it demands a well-defined channel model that captures relativistic propagation effects, complex background noise, and cosmic structures that can distort or attenuate the waves.

\subsubsection{Implications for Communication System Design}
The unresolved challenges in gravitational communication have direct implications for system design:
\begin{itemize}
    \item \textbf{Signal Design and Spectrum Planning:} Signals must be designed to account for frequency shifts due to gravitational redshift and to avoid frequency bands with high attenuation or noise. For example, frequency bands affected by superradiance near rotating black holes may need to be excluded or compensated for \cite{fier2021}. Effective spectrum planning is critical to optimize signal clarity and reduce interference risks.
    \item \textbf{Receiver Design:} Receivers must handle complex propagation effects, including polarization changes and multipath distortions. Advanced filtering techniques are necessary to mitigate noise and extract the signal effectively, especially in environments with high background noise or coherent interference.
    \item \textbf{Signal Processing Techniques:} Compensation algorithms are essential to correct for phase distortions and frequency shifts caused by cosmic structures. Additionally, noise reduction methods are needed to enhance the signal-to-noise ratio (SNR) for clearer detection. These techniques should be adaptable to varying propagation conditions and cosmic environments.
\end{itemize}

\subsubsection{Need for Groundbreaking Theoretical and Experimental Research}
Advancing our understanding of these open issues requires a combination of theoretical innovations and experimental efforts:
\begin{itemize}
    \item \textbf{Developing Quantum Gravity Theories:} Formulating and testing unified theories, such as string theory or loop quantum gravity, is critical to understanding gravitational wave behavior at quantum scales.
    \item \textbf{Exploring Quantum Effects on Gravitational Waves:} Designing high-sensitivity experiments to detect potential quantum signatures in gravitational wave signals, possibly using advanced detectors, could uncover new phenomena.
    \item \textbf{Simulating Quantum-Gravitational Interactions:} High-performance computing can be used to model and simulate the interactions between gravitational waves and quantum fields, offering insights into their combined effects on signal propagation.
\end{itemize}

By addressing these research challenges, the scientific community can work towards designing more robust communication systems that account for both classical and quantum influences on gravitational wave propagation.

After thoroughly investigating these open issues and their interrelationships, we can enhance the reliability and effectiveness of gravitational communication across various cosmic environments. Such efforts will enable the development of advanced communication systems capable of operating under diverse and challenging conditions, paving the way for future applications in deep-space communication and cosmic exploration.

\section{Gravitational vs. Classical Communications}

Building on the advancements in gravitational communication discussed in previous sections—particularly in signal generation, detection, modulation, and channel analysis—it is essential to compare this emerging technology with established communication methods. Gravitational wave generation in lab condition involves generating extremely weak signals, using methods like high-powered lasers, electromagnetic fields, and twisted light beams. Detection relies on highly sensitive instruments such as laser interferometers and advanced data processing techniques, including direct and indirect approaches. Modulation techniques remain largely theoretical, facing significant technical barriers. Gravitational waves have the unique ability to penetrate dense cosmic environments without significant attenuation, making them theoretically advantageous for deep space communication.

The following subsections provide detailed comparisons between gravitational communication and specific conventional communication technologies, focusing on signal generation, detection, modulation, and transmission characteristics, while highlighting key differences without repeating the details of gravitational communication.

\subsection{Gravitational vs. Electromagnetic Communication}

EM communication employs mature technologies for generating and detecting signals across a wide frequency spectrum, from low-frequency radio waves to high-frequency millimeter waves used in 5G and 6G networks \cite{pereira2022}. Electromagnetic waves do not require a medium to propagate and travel at the speed of light, making them suitable for space-based communication within our solar system and nearby cosmic regions. The versatility of EM waves is evident in their ease of generation, modulation, and detection using well-established techniques. This makes them indispensable for various applications, including satellite communication, space exploration, and astronomical observations. However, EM waves interact strongly with matter and electromagnetic fields. When passing through media such as interstellar gas, cosmic dust, or plasma, EM signals can experience interference, attenuation, and absorption. These interactions can scatter and weaken the signal, limiting the effectiveness of EM communication in dense or highly magnetized regions of space.

In contrast, gravitational waves are disturbances in spacetime itself, arising from massive accelerating objects such as merging black holes or neutron stars. Unlike EM waves, gravitational waves are not affected by electromagnetic fields or most types of matter. This allows them to pass through cosmic obstacles, such as interstellar dust clouds or even entire stars, with minimal attenuation. While both EM and gravitational waves travel at the speed of light, gravitational waves maintain their amplitude and phase across long distances and dense environments. They are thus well-suited for potential interstellar communication, especially in regions where EM waves may suffer signal degradation \cite{akyildiz2004}. An additional advantage of gravitational waves is their inherent physical layer security. Due to current technological limitations, gravitational waves are highly inaccessible and cannot be easily intercepted or tampered with. This makes them resistant to eavesdropping and unauthorized access, enhancing the security of transmissions. Unlike EM waves, which can be relatively easily intercepted and decoded by unintended recipients, gravitational wave signals are challenging to detect without highly specialized equipment.

The relationship between EM and gravitational communication lies in their complementary properties. EM communication is highly effective for near-Earth and short-to-medium-range space communication due to its ease of use and the mature infrastructure supporting it. It handles local and regional data transmission efficiently. Gravitational communication, while still theoretical in practical application, offers a promising alternative for stable, long-range communication in extreme or densely populated cosmic environments. In scenarios where EM waves are hindered by cosmic interference or security concerns, gravitational waves could provide a reliable and secure communication channel.

Moreover, advancements in gravitational wave detection technology could benefit from EM communication systems. For example, coordinating a network of gravitational wave detectors spread across vast distances would require efficient EM communication to synchronize observations and share data in real-time.

\subsection{Gravitational vs. Quantum Communication}

Quantum communication leverages the principles of quantum mechanics to achieve secure data transmission, primarily through quantum key distribution (QKD) \cite{mehic2020}. Quantum states are highly sensitive to environmental disturbances, making them susceptible to decoherence from factors like temperature fluctuations and electromagnetic noise. This fragility requires that quantum communication systems be carefully shielded or even operated at cryogenic temperatures to preserve signal integrity, especially over long distances. While quantum entanglement enables an instantaneous correlation between particles, it does not permit faster-than-light communication. Quantum communication also faces challenges in transmitting information over large interstellar distances, as maintaining entanglement and coherence in the quantum states can be difficult in the vast and varied cosmic environment.

Gravitational waves, by contrast, do not rely on quantum states or entangled particles. They are disturbances in spacetime itself, which means they are inherently stable and unaffected by electromagnetic noise or environmental interference. Unlike quantum states, which can be disrupted by the surrounding environment, gravitational waves maintain their integrity through even extreme conditions, such as dense interstellar clouds or strong magnetic fields.

The interplay between gravitational wave and quantum communication centers on the trade-off between security and practicality over long distances. Quantum communication offers unparalleled security features based on quantum mechanical principles but is limited in range and sensitive to environmental noise, especially in space. Gravitational communication provides a robust alternative for long-range communication with inherent physical layer security. Its resilience to interference and resistance to eavesdropping suggest potential for secure messaging across interstellar distances. Additionally, research into quantum gravity might reveal new ways to encode information into gravitational waves using quantum states, potentially combining the security of quantum communication with the reliability and inherent security of gravitational waves.

\subsection{Gravitational vs. Particle-Based Communication}
Particle-based communication refers to the use of particles at the subatomic or molecular level to accomplish the process of encoding, transmitting and decoding information. Unlike traditional communication methods that rely on the propagation of electromagnetic waves or photons, particle-based communication focuses on the use of a variety of particles, such as neutrinos, molecules, and ions, as information carriers. The advantages or limitations usually depend on the nature of the particles chosen and how they interact with their surroundings. For example, the use of neutrinos, which have a very small mass and interact weakly with matter, can realize ultra-strong penetration and long-distance transmission; whereas molecular communication based on molecular diffusion or flow is suitable for micro-scale, short-distance and precise information exchange \cite{akan2016}. Different forms of particle communication are often suitable for different application situations, which may be applied to macroscopic and long-distance cosmic communication \cite{aartsen2017}, as well as high-precision and biologically friendly communication in microscopic and localized environments \cite{xiao2023}.

\begin{table*}[h!]
\centering
\caption{Comparison of Different Communication Methods}
\begin{tabular}{|>{\raggedright\arraybackslash}p{2cm}|
                  >{\raggedright\arraybackslash}p{2.5cm}|
                  >{\raggedright\arraybackslash}p{2.5cm}|
                  >{\raggedright\arraybackslash}p{3.5cm}|
                  >{\raggedright\arraybackslash}p{2.5cm}|
                  >{\raggedright\arraybackslash}p{2.5cm}|}

\hline
\textbf{Communication Method} & \textbf{Energy Budget} & \textbf{Range}       & \textbf{Rate}                                                                       & \textbf{Medium}                     & \textbf{Anti-Interference Ability} \\ \hline
\textit{\textbf{\color{red}Gravitational Communication}} & Very high, hard to generate and requires astronomical events like black hole mergers in nature & Cosmic scale, can penetrate stars and dense matter & Very low communication rate (Hz to kHz bandwidth, narrowband); propagation speed equals the speed of light & No medium required & Very strong, almost immune to environmental interference \\ \hline
\textit{Electromagnetic Communication} & Reasonable, can be optimized by directional emission & Interstellar range, affected by interstellar medium (e.g., absorption, scattering) & Very high communication rate (Gbps to Tbps); propagation speed equals the speed of light & No medium required & Moderate, prone to electromagnetic interference \\ \hline
\textit{Quantum Communication}   & Low energy consumption for a single transmission, but high overall energy consumption of the equipment & Fiber: hundreds of kilometers; Free-space: thousands of kilometers; limited for interstellar & Moderate communication rate (currently at Mbps in experiments); propagation speed is near the speed of light via photons & Requires fiber or free-space, depends on high-precision quantum devices & Weak, highly sensitive to environmental interference \\ \hline
\textit{Neutrino Communication}   & Very high, requires large particle accelerators to produce high-energy neutrino beams  & Cosmic scale, can penetrate stars and dense matter & Extremely low communication rate (only a few bits per second in experiments); propagation speed close to the speed of light in vacuum & No medium required & Very strong, unaffected by environmental interference \\ \hline
\textit{Molecular Communication}  & Low, suitable for micro-scale biological systems & Very limited, usually from micrometers to millimeters & Very low communication rate (typically in bps); propagation speed depends on diffusion or flow, much slower than the speed of sound & Requires medium (e.g., liquid, gas), relies on diffusion or flow & Weak, highly influenced by environments \\ \hline
\textit{Acoustic Communication}  & Low, suitable for low-power devices & Limited, from kilometers to tens of kilometers; cannot propagate in vacuum & Low communication rate (typically in kbps); propagation speed is the speed of sound (approx. 340 m/s, highly medium-dependent) & Requires medium (e.g., gas, liquid, solid) & Weak, highly affected by noise and medium variability \\ \hline
\textit{Optical Communication}  & Reasonable, efficient directional transmission & Interstellar range, but affected by interstellar dust and atmospheric absorption & Very high communication rate (commonly Gbps to Tbps); propagation speed equals the speed of light & Requires free space beyond the atmosphere or optical fiber & Moderate, affected by light scattering and absorption \\ \hline
\end{tabular}
\label{tab:communication_methods}
\end{table*}

Neutrino communication utilizes neutrinos, nearly massless particles with very weak interactions with matter, to transmit information. This weak interaction makes neutrinos extremely penetrative, allowing them to pass through dense objects like planets or stars with minimal attenuation. However, due to this low interaction probability, producing and detecting neutrinos require high-energy sources, such as particle accelerators or nuclear reactors, and large-scale, sensitive detectors, often positioned deep underground or within ice to avoid background noise \cite{aartsen2017}. These requirements make it challenging to achieve strong and stable neutrino signals without substantial infrastructure, also result in low signal intensity with high error rates.

Another particle-based communication method is molecular communication, which uses molecules as information carriers. It relies on diffusion or flow mechanisms through liquid or gas media to transmit signals. The properties of molecular communication make it ideally suited for short-range applications at the micrometer to millimeter level, such as cell-to-cell biological signaling \cite{kuscu2019}. However, since propagation relies on diffusion, it is extremely slow, often below the speed of sound; at the same time, molecular communication is highly sensitive to the nature of the medium and is strongly influenced by temperature, medium density and ambient noise. These factors make it difficult to meet large-scale or high-precision communication needs.

By contrast, gravitational waves generated by large-scale cosmic events may carry higher energy signals over vast distances, maintaining signal strength more effectively than neutrinos. As spacetime disturbances, gravitational waves are not limited by particle interactions and can retain their amplitude and phase over cosmological scales, providing a more stable signal for long-range communication. This characteristic makes them uniquely suited for interstellar communication. Unlike molecular communication, gravitational waves are entirely independent of medium, allowing seamlessly through diverse environments, including the vacuum of space, without any attenuation. Furthermore, gravitational waves are unaffected by environmental factors such as temperature or density and can penetrate high-density or complex regions, such as stellar interiors or interstellar dust clouds, with minimal interference. Finally, their propagation speed reaches the speed of light, significantly outpacing the diffusion-dependent speed of molecular communication and enabling the potential for rapid, efficient communication across cosmic distances.

Particle-based communication methods, guided by different principles, find diverse applications in distinct contexts. Neutrino communication, similar to gravitational waves, is suited for macroscopic, long-distance scenarios due to its penetrative and resilient properties. In contrast, molecular communication, fundamentally different from gravitational waves, excels in microscopic, short-distance applications where precision and low energy consumption are crucial. These various approaches provide possibilities for communication across both vast interstellar scales and localized biological systems. Together, they could inspire new frameworks for achieving the connectivity of everything in the future.

\subsection{Gravitational vs. Acoustic Communication}

Acoustic communication relies on sound waves, which are longitudinal waves. In longitudinal waves, particle displacement occurs in the same direction as the wave’s propagation, meaning that the particles in the medium oscillate back and forth along the direction of travel. This particle-based vibration requires a material medium, such as air, water, or a solid substance, for transmission. Consequently, sound waves cannot travel through a vacuum and are confined to environments where a suitable medium exists, like underwater or within Earth's atmosphere. Additionally, sound waves are relatively slow compared to light-speed waves, with speeds that depend on the medium (approximately 340 m/s in air, 1500 m/s in water, and faster in denser solids) \cite{truax2001}.

Gravitational waves, on the other hand, are transverse waves—ripples in spacetime itself, generated by massive accelerating bodies. In transverse waves, the oscillation occurs perpendicular to the direction of wave propagation. This perpendicular, spacetime-based disturbance allows gravitational waves to propagate through a vacuum without needing any medium, making them uniquely suited for interstellar communication. Additionally, unlike sound waves, gravitational waves travel at the speed of light ($3 \times 10^8 \, \text{m/s}$) and can maintain their integrity over vast cosmic distances, penetrating dense cosmic structures such as gas clouds and even stars with minimal attenuation.

While acoustic communication is practical for short-range transmission within confined environments on Earth, it lacks any applicability in the vacuum of space. Gravitational waves, with their transverse nature, ability to travel through both vacuum and dense matter, broad frequency range, and resilience against environmental interference, offer a distinct advantage for potential long-distance interstellar communication. Their independence from a transmission medium and their robustness in extreme environments make gravitational waves far more viable for cosmic-scale messaging where sound waves are entirely unusable.

\subsection{Gravitational vs. Optical Communication}

Optical communication uses light waves, specifically electromagnetic waves within the visible spectrum, which are transverse waves that propagate through oscillating electric and magnetic fields \cite{kaushal2016}. While capable of traveling through a vacuum, free-space optical communication can be affected by atmospheric conditions like fog, rain, and dust, and requires a clear line-of-sight path. Optical signals can also be scattered or absorbed by interstellar media, limiting their effectiveness for deep space communication. Optical communication is discussed separately because, although both optical and quantum communication rely on photons as information carriers, they differ fundamentally. Optical communication is rooted in classical electromagnetic principles and focuses on maximizing data rates and transmission efficiency, while quantum communication relies on inherently different mechanisms like quantum superposition and entanglement, prioritizing security over distance and speed.

Unlike light, gravitational waves are largely immune to interference from matter or electromagnetic fields, can propagate through any medium without attenuation and without the need for line-of-sight. This resilience makes gravitational waves highly suitable in regions where light waves may be scattered or absorbed, presenting a significant theoretical advantage. While optical communication is effective for many applications, especially within Earth's atmosphere and for certain space missions, its limitations highlight the challenges faced in interstellar communication.

Gravitational and optical communications could potentially complement each other in a multi-modal communication network. Optical communication could be used where conditions are favorable, leveraging its high data rates, while gravitational communication could provide reliable links where optical signals are obstructed or degraded. Besides, advances in optical technologies, such as lasers and photodetectors, play a crucial role in gravitational wave detection systems like LIGO, which use laser interferometry to measure spacetime distortions. Improvements in optical components could enhance the sensitivity and capabilities of gravitational wave detectors.

\section{Advantages, Applications, and Future Vision of Gravitational Wave Communication}

 \begin{figure*}[!t]
    \centering
    \includegraphics[width=\linewidth]{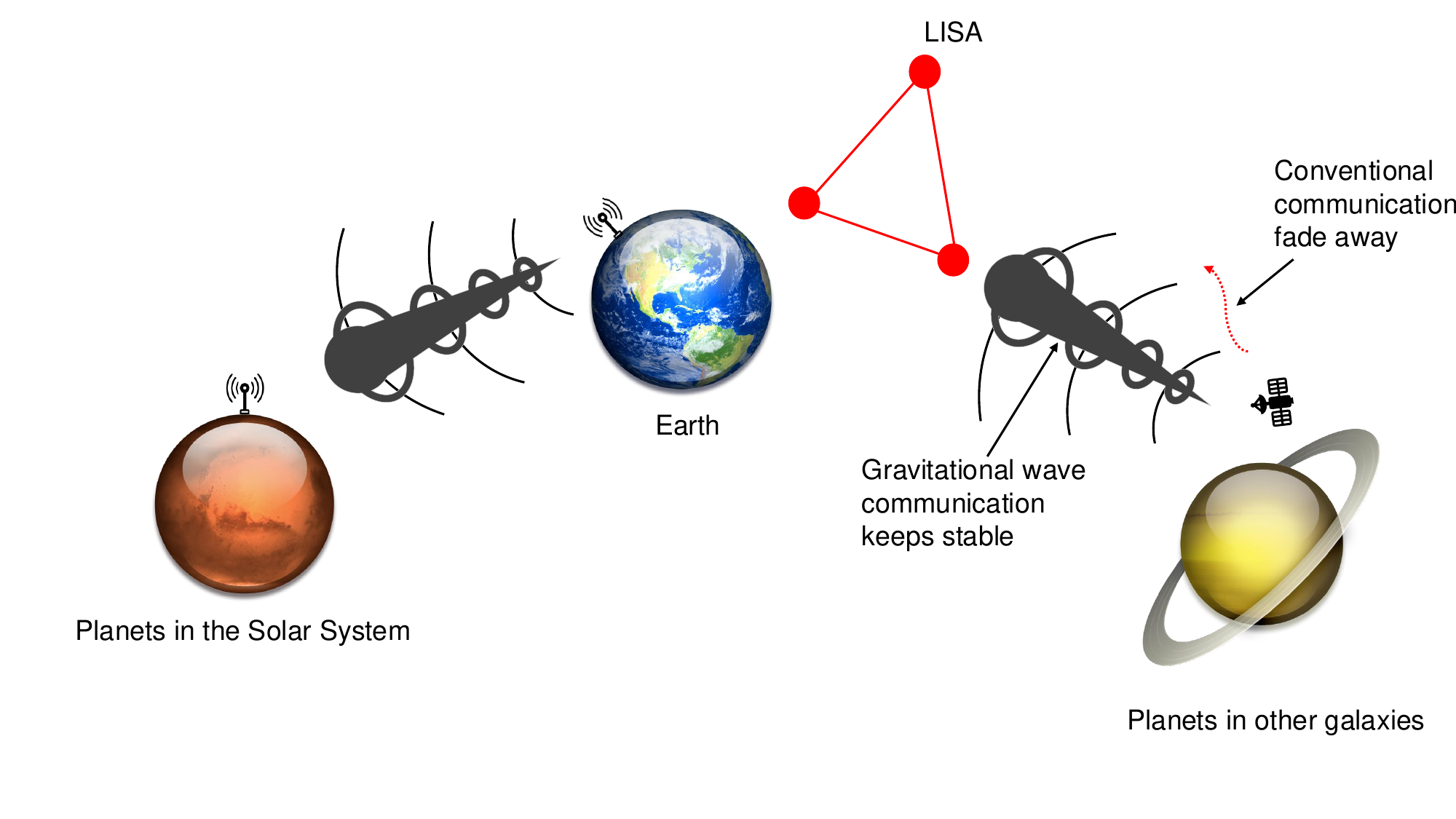}
    \caption{Gravitational wave used for interstellar communication.}
    \label{fig:example}
\end{figure*}

Gravitational communication stands out as a transformative technology, fundamentally distinct from traditional communication methods. By leveraging the unique properties of spacetime distortions, it offers unparalleled advantages, particularly in interstellar and extreme environments. Its key benefits, potential applications, and long-term vision are best understood through a comparative analysis with conventional methods. As shown in Table~\ref{tab:communication_methods}, various communication techniques are compared across key parameters such as energy budget, range, data rate, required medium, and anti-interference ability. This consolidated comparison highlights the unique strengths of gravitational wave communication, illustrating its potential to overcome the limitations of traditional systems and paving the way for groundbreaking advancements in extreme communication scenarios.

\subsection{Unique Advantages of Gravitational Wave Communication}

Gravitational waves are fundamentally different from electromagnetic waves, enabling them to overcome many limitations of conventional communication technologies. The key advantages include:

\subsubsection{Robustness in Extreme Environments} Gravitational waves remain largely unaffected by material obstructions or electromagnetic interference, allowing them to propagate through dense cosmic structures such as black holes, interstellar gas clouds, and dust with minimal attenuation. This resilience ensures high signal fidelity even across vast interstellar distances.

\subsubsection{Minimal Energy Loss Over Long Distances} Unlike electromagnetic waves, which dissipate energy due to interactions with intervening media, gravitational waves experience negligible energy loss during propagation. This property makes them ideal for long-range communication where energy efficiency is critical.

\subsubsection{Immunity to Common Communication Challenges} Issues like scattering, diffraction, and reflection that plague traditional communication systems are virtually nonexistent for gravitational waves, ensuring more reliable and stable data transmission.

\subsubsection{Potential for Synergy with Natural Phenomena} The ability to harness naturally occurring gravitational waves from cosmic events can reduce energy requirements for communication systems, leveraging the universe itself as a medium for data transfer.

\subsection{Applications in Deep Space Exploration}

Deep space exploration presents unique communication challenges, such as vast distances, interference from cosmic phenomena, and limited relay infrastructure.  As illustrated in Fig.~\ref{fig:example}, gravitational communication offers a promising solution to these challenges:

\begin{itemize}
    \item \textbf{Reliable Interstellar Communication:} Gravitational waves can maintain consistent signal quality over immense distances, making them suitable for missions beyond the solar system. For instance, communication with spacecraft exploring nearby star systems or exoplanets could rely on gravitational waves, eliminating the need for error-prone amplification or relay systems.

    \item \textbf{Resilience Against Environmental Interference:} Unlike radio frequencies used in Mars-to-Earth communications, which are vulnerable to disruptions from solar activity or planetary dust storms \cite{civas2021}, \cite{akyildiz2003}, gravitational waves bypass such challenges. This ensures uninterrupted data transfer, even during critical mission phases.

    \item \textbf{Scalability for Long-Distance Networks:} By integrating gravitational wave detectors into a distributed network, interstellar communication systems can achieve greater coverage and reliability, enabling large-scale exploratory missions with robust communication frameworks.

\end{itemize}

\subsection{Harnessing Natural Astrophysical Phenomena for Communication}

Astrophysical events such as binary star mergers, neutron star collisions, and supernovae produce significant gravitational waves. By leveraging these naturally occurring phenomena, it becomes possible to create energy-efficient communication systems:

\begin{itemize}
    \item \textbf{Amplification Through Natural Sources:} Gravitational communication systems can ``piggyback" on waves generated by high-energy cosmic events, reducing the need for artificial signal generation. This approach minimizes energy consumption while enabling long-distance data exchange.

    \item \textbf{Interstellar Network Integration:} A network of gravitational wave detectors positioned across interstellar distances could synchronize with both artificial and natural gravitational waves, enabling unprecedented levels of data exchange on a cosmic scale.

\end{itemize}
\subsection{Overcoming Challenges in Terrestrial and Subsurface Environments}

Gravitational communication is not restricted to space applications; it also offers solutions to terrestrial and subsurface challenges where traditional methods fail:

\begin{itemize}
    \item \textbf{High-Energy Plasma and Fusion Environments:} Communication in environments such as fusion reactors, where electromagnetic signals degrade rapidly, can benefit from the penetration capabilities of gravitational waves.

    \item \textbf{Geological and Subsurface Communication:} Dense geological formations, often encountered in mining or geophysical research, attenuate conventional signals. Gravitational waves, however, maintain their integrity, enabling reliable subsurface communication systems.

    \item \textbf{Disaster Response and Emergency Communication:} Natural disasters like earthquakes or volcanic eruptions often disrupt conventional networks. A gravitational communication system could remain operational in such scenarios, providing critical links for rescue and coordination efforts.

\end{itemize}

\subsection{Future Prospects and Vision}

Gravitational wave communication represents a paradigm shift, opening new avenues for connectivity across diverse domains:

\begin{itemize}
    \item \textbf{Interdisciplinary Research Directions:} Advancing gravitational wave communication requires breakthroughs in modulation techniques, efficient detection technologies, and cost-effective wave generation. Collaborative efforts across physics, engineering, and materials science are crucial to realizing its potential.

    \item \textbf{Integration with Existing Communication Systems:} Combining gravitational communication with electromagnetic and quantum networks can create multi-modal frameworks, enhancing the robustness and versatility of global communication systems.

    \item \textbf{Redefining Cosmic Connectivity:} Gravitational communication has the potential to redefine our approach to interstellar connectivity, enabling secure, energy-efficient, and reliable information exchange across astronomical distances.
\end{itemize}

Gravitational communication’s unparalleled resilience, combined with its diverse applications in deep space exploration, terrestrial environments, and emergency scenarios, positions it as a groundbreaking frontier in communication technology. Continued research and development in this field will not only address current challenges but also unlock new possibilities for connectivity, making gravitational communication a cornerstone of humanity's efforts to bridge distances both on Earth and across the stars.

\section{Conclusions}
This paper systematically examines the current concepts and methodologies in gravitational communication, aiming to serve as a reference for the development of this emerging technology. Gravitational communication, as a frontier research direction with significant potential, is gradually moving from theoretical exploration to practical application. Despite challenges such as weak signal strength, low detection sensitivity, and environmental interference, sustained research has provided a strong foundation for addressing these issues. In addition to reviewing advancements, this paper also presents a comparative analysis between gravitational communication and other established methods—including electromagnetic, optical, acoustic, quantum, and particle communications—highlighting the distinct strengths and limitations of each in terms of generation, detection, modulation, and transmission. While these conventional communication technologies have matured and are widely utilized, gravitational waves offer unique advantages, particularly for long-distance, interference-free communication in unconventional environments. This comparison underscores the gaps that remain in making gravitational communication practical but also emphasizes the complementary potential alongside existing technologies. Although a fully practical gravitational wave communication system remains unfeasible, we aim to use this survey to highlight its potential and stimulate further research and innovation, especially for space communication scenario.



\ifCLASSOPTIONcaptionsoff
  \newpage
\fi



%


\bibliographystyle{IEEEtran}
\bibliography{references}

\vfill

\end{document}